\documentclass[twocolumn,showkeys,preprintnumbers,aps,a4paper,amssymb,prd,superscriptaddress,nofootinbib]{revtex4-2}
\usepackage{comment}
\usepackage{graphicx}
\usepackage{epsf}
\usepackage{bm}
\usepackage{amsmath}
\usepackage{amsfonts}
\usepackage{amssymb}
\usepackage{epstopdf}
\usepackage{natbib}
\usepackage{color}
\usepackage[dvipsnames]{xcolor}
\usepackage{verbatim}
\usepackage{multirow}
\usepackage{soul}
\usepackage{physics}
\usepackage{bm}
\usepackage{microtype}
\usepackage[colorlinks = true,
            linkcolor = blue,
            urlcolor  = blue,
            citecolor = blue,
            anchorcolor = blue]{hyperref}

\usepackage{amsmath}
\usepackage[capitalize]{cleveref}
\usepackage[normalem]{ulem}
\usepackage{enumitem}
\usepackage{booktabs}

\usepackage{lipsum}

\makeatletter\let\expandableinput\@@input\makeatother

\definecolor{brandeisblue}{rgb}{0.0, 0.44, 1.0}

\definecolor{brightpink}{rgb}{1.0, 0.0, 0.5}
\definecolor{lava}{rgb}{0.81, 0.06, 0.13}
\definecolor{lavenderindigo}{rgb}{0.58, 0.34, 0.92}

\def\Mpl{M_{\rm P}}

\begin{document}

\preprint{YITP-24-57}

\title{Cosmological constraints on $\Lambda_{\rm s}$CDM scenario in a type II minimally modified gravity}

\author{\"{O}zg\"{u}r Akarsu}
\email{akarsuo@itu.edu.tr}
\affiliation{Department of Physics, Istanbul Technical University, Maslak 34469 Istanbul, T\"{u}rkiye}

\author{Antonio De Felice}
\email{antonio.defelice@yukawa.kyoto-u.ac.jp}
\affiliation{
Center for Gravitational Physics and Quantum Information, Yukawa Institute for Theoretical Physics, Kyoto University, 606-8502, Kyoto, Japan
}

\author{Eleonora Di Valentino}
\email{e.divalentino@sheffield.ac.uk}
\affiliation{School of Mathematics and Statistics, University of Sheffield, Hounsfield Road, Sheffield S3 7RH, United Kingdom}

\author{Suresh Kumar}
\email{suresh.kumar@plaksha.edu.in}
\affiliation{Data Science Institute, Plaksha University, Mohali, Punjab-140306, India}

\author{Rafael C. Nunes}
\email{rafadcnunes@gmail.com}
\affiliation{Instituto de F\'{i}sica, Universidade Federal do Rio Grande do Sul, 91501-970 Porto Alegre RS, Brazil}
\affiliation{Divis\~ao de Astrof\'isica, Instituto Nacional de Pesquisas Espaciais, Avenida dos Astronautas 1758, S\~ao Jos\'e dos Campos, 12227-010, SP, Brazil}

\author{Emre \"{O}z\"{u}lker}
\email{ozulker17@itu.edu.tr}
\affiliation{Department of Physics, Istanbul Technical University, Maslak 34469 Istanbul, T\"{u}rkiye}
\affiliation{School of Mathematics and Statistics, University of Sheffield, Hounsfield Road, Sheffield S3 7RH, United Kingdom}

\author{J. Alberto Vazquez}
\email{javazquez@icf.unam.mx}
\affiliation{Instituto de Ciencias F\'isicas, Universidad Nacional Aut\'onoma de M\'exico, Cuernavaca, Morelos, 62210, M\'exico}

\author{Anita Yadav}
\email{anita.math.rs@igu.ac.in}
\affiliation{Department of Mathematics, Indira Gandhi University, Meerpur, Haryana 122502, India}

\begin{abstract}
The idea of a rapid sign-switching cosmological constant (mirror AdS-dS transition) in the late universe at $z\sim1.7$, known as the $\Lambda_{\rm s}$CDM model, has significantly improved the fit to observational data and provides a promising scenario for alleviating major cosmological tensions, such as the $H_0$ and $S_8$ tensions. However, in the absence of a fully predictive model, implementing this fit required conjecturing that the dynamics of the linear perturbations are governed by general relativity. Recent work embedding the $\Lambda_{\rm s}$CDM model with the Lagrangian of a type II minimally modified gravity known as VCDM has propelled $\Lambda_{\rm s}$CDM to a fully predictive model, removing the uncertainty related to the aforementioned assumption; we call this new model $\Lambda_{\rm s}$VCDM. In this work, we demonstrate that not only does $\Lambda_{\rm s}$CDM fit the data better than the standard $\Lambda$CDM model, but the new model, $\Lambda_{\rm s}$VCDM, performs even better in alleviating cosmological tensions while also providing a better fit to the data, including cosmic microwave background, baryon acoustic oscillations, type Ia supernovae, and cosmic shear measurements. Our findings highlight the $\Lambda_{\rm s}$CDM framework, particularly the $\Lambda_{\rm s}$VCDM model, as a compelling alternative to the standard $\Lambda$CDM model, especially by successfully alleviating the $H_0$ tension. Additionally, these models predict higher values for $\sigma_8$, indicating enhanced structuring, albeit with lower present-day matter density parameter values and consequently reduced $S_8$ values, alleviating the $S_8$ tension as well. This demonstrates that the data are well fit by a combination of background and linear perturbations, both having dynamics differing from those of $\Lambda$CDM. This paves the way for further exploration of new ways for embedding the sign-switching cosmological constant into other models.
\end{abstract}


\maketitle
%

\section{Introduction}

The standard $\Lambda$CDM model has been remarkably consistent with the majority of data from astrophysical and cosmological observations conducted over the past decades~\cite{Planck:2018vyg,eBOSS:2020yzd,ACT:2020gnv,KiDS:2020suj,SPT-3G:2022hvq,DES:2024tys,DESI:2024mwx}. However, in the new era of high-precision cosmology, certain discrepancies, such as the $H_0$ tension~\cite{Verde:2019ivm,DiValentino:2020zio} reaching to $5\sigma$ level of significance~\cite{Riess:2021jrx,Breuval:2024lsv,Murakami:2023xuy} and the $S_8$ tension reaching $3\sigma$~\cite{DiValentino:2020vvd,KiDS:2020suj,Chen:2021wdi,Burger:2023qef,Amon:2022azi,Dalal:2023olq,Kilo-DegreeSurvey:2023gfr,Li:2023azi,Harnois-Deraps:2024ucb}, along with some others of lesser significance, have emerged when analyzing different datasets, bringing the standard model to a crossroads. This pivotal situation has compelled the scientific community to embark on a quest for alternative explanations, either rooted in novel physics or through the identification of potential systematic errors in the data. For recent reviews, see Refs.~\cite{DiValentino:2021izs,Abdalla:2022yfr,Perivolaropoulos:2021jda,DiValentino:2022fjm,Kamionkowski:2022pkx,Vagnozzi:2023nrq,Akarsu:2024qiq,Khalife:2023qbu}.

Along the path of seeking novel physics as an explanation for cosmological tensions, many attempts have been made; see, e.g., Refs.~\cite{DiValentino:2021izs,Perivolaropoulos:2021jda,Abdalla:2022yfr} and references therein for a comprehensive but not exhaustive list. Most of these rely on a bottom-up approach, assuming the existence of some effective field theory that could account for the phenomenology assumed. In this approach, distinguishing between the many possible models can be achieved by identifying which model best fits the observational data. In this sense, one of the most promising ideas is the recently proposed $\Lambda_{\rm s}$CDM theory-framework, which considers the possibility that the Universe has recently (at redshift $z \sim 2$) undergone a rapid \textit{mirror} anti--de Sitter (AdS) vacuum to a de Sitter (dS) vacuum transition (a sign-switching of the cosmological constant, $\Lambda_{\rm s}$)~\cite{Akarsu:2019hmw, Akarsu:2021fol, Akarsu:2022typ, Akarsu:2023mfb,Yadav:2024duq}. This simple paradigm, standing as one of the most economical approaches (introducing only one additional parameter on top of $\Lambda$CDM, $z_\dagger$, the redshift of the AdS-dS transition) for the simultaneous resolution of major cosmological tensions in the literature so far, can indeed account for a plethora of different datasets, as we will also see in this work, and it attracts interest from both theoretical and observational points of view. The suggested rapid nature of the sign-switching cosmological constant, along with its shift from negative to positive values, has generally found challenging in identifying a concrete physical mechanism. However, the phenomenological success of $\Lambda_{\rm s}$CDM, despite its simplicity, has led to increasing interest in introducing theoretical approaches for the realization of the late-time mirror AdS-dS(-like) transition. It was shown in~\cite{Anchordoqui:2023woo,Anchordoqui:2024gfa,Antoniadis:2024sfa} that although the AdS swampland conjecture suggests that AdS-dS transition in the late universe seems unlikely (due to the arbitrarily large distance between AdS and dS vacua in moduli space), it can be realized through the Casimir forces of fields inhabiting the bulk. Furthermore, it was demonstrated in~\cite{Alexandre:2023nmh} that in various formulations of GR, it is possible to obtain a sign-switching cosmological constant  through an overall sign change of the metric. Recently, in~\cite{Akarsu:2024qsi}, the authors proposed embedding the late time mirror AdS-dS transition into the theoretical framework of VCDM~\cite{DeFelice:2020eju,DeFelice:2020cpt,DeFelice:2020onz,DeFelice:2020prd,DeFelice:2021xps,DeFelice:2022uxv}, a minimal theory of gravity, i.e., a model that does not introduce extra degrees of freedom into the theory. The result of this embedding leads to a theory that we will call the $\Lambda_{\rm s}$VCDM model. In this way, $\Lambda_{\rm s}$CDM has become a fully predictive model with the ability to describe all gravitational phenomena, including the cosmological evolution of our Universe. We refer readers to Refs.~\cite{Dutta:2018vmq,Visinelli:2019qqu,Perez:2020cwa,DiValentino:2020naf,Acquaviva:2021jov,Sen:2021wld,Ozulker:2022slu,DiGennaro:2022ykp,Moshafi:2022mva,vandeVenn:2022gvl,Ong:2022wrs,Tiwari:2023jle,Vazquez:2023kyx,Adil:2023exv,Adil:2023ara,Paraskevas:2023itu,Wen:2023wes,Adil:2023exv,Menci:2024rbq,Gomez-Valent:2024tdb,Felice_2024,Manoharan:2024thb} for more works that study dark energy assuming negative density values, (mostly) consistent with a negative (AdS-like) cosmological constant, for $z \gtrsim 1.5-2$, particularly aiming to address cosmological tensions such as the $H_0$ and $S_8$ tensions and, recently, anomalies from JWST, and to Refs.~\cite{Sahni:2014ooa,BOSS:2014hhw,Poulin:2018zxs,Wang:2018fng,Calderon:2020hoc,Bonilla:2020wbn,Escamilla:2021uoj,Bernardo:2021cxi,Akarsu:2022lhx,Bernardo:2022pyz,Malekjani:2023ple,Escamilla:2023shf,Gomez-Valent:2023uof,Medel-Esquivel:2023nov,DESI:2024aqx,Bousis:2024rnb,Wang:2024hwd,Colgain:2024ksa,Sabogal:2024qxs} suggesting such dynamics for dark energy from model-independent/nonparametric observational reconstructions and investigations.

In this paper, we will explore how implementing $\Lambda_{\rm s}$CDM into a model affects cosmological observables, particularly concerning the $H_0$~\cite{Riess:2021jrx,Verde:2019ivm,Breuval:2024lsv,DiValentino:2021izs} and $S_8$~\cite{DiValentino:2020vvd,Harnois-Deraps:2024ucb} tensions. The $\Lambda_{\rm s}$CDM model, considered within the framework of general relativity, alters the background dynamics compared to $\Lambda$CDM without modifying the equations of motion for the perturbations. In contrast, the $\Lambda_{\rm s}$VCDM model has a well-defined Lagrangian, which generally leads to differences from $\Lambda$CDM in both the background and perturbation equations of motion. Specifically, $\Lambda_{\rm s}$VCDM is defined as the model introduced in~\cite{Akarsu:2024qsi}, sharing the same background as $\Lambda_{\rm s}$CDM but with one additional parameter compared to $\Lambda$CDM: the redshift at which the transition occurs.

Since the observables we consider depend on both the background and cosmological linear perturbation dynamics, it is expected that $\Lambda_{\rm s}$CDM and $\Lambda_{\rm s}$VCDM will, in general, yield different constraints from the data. Even if $\Lambda_{\rm s}$CDM provides a good fit to the data (compared to $\Lambda$CDM), it is not clear a priori whether a full model implementation of $\Lambda_{\rm s}$CDM, namely $\Lambda_{\rm s}$VCDM, will continue to provide a good fit to the data. Therefore, to address this uncertainty, we will compare $\Lambda$CDM, $\Lambda_{\rm s}$CDM, and $\Lambda_{\rm s}$VCDM using the same datasets in this paper.

The paper is organized as follows. In~\cref{sec:models}, we present the scenarios explored in this work, namely the $\Lambda_{\rm s}$CDM and $\Lambda_{\rm s}$VCDM. In~\cref{sec:datasets}, we outline the datasets and the methodology used to analyze these scenarios. In~\cref{sec:results}, we discuss the results obtained. Finally, in~\cref{sec:conclusions}, we derive our conclusions.

\section{$\Lambda_{\rm s}$CDM paradigm and its implementation into the VCDM model} \label{sec:models}

The $\Lambda_{\rm s}$CDM paradigm is inspired by the recent conjecture that the universe underwent a spontaneous mirror AdS-dS transition characterized by a sign-switching cosmological constant ($\Lambda_{\rm s}$) around $z \sim 2$~\cite{Akarsu:2019hmw, Akarsu:2021fol, Akarsu:2022typ, Akarsu:2023mfb, Yadav:2024duq}. This conjecture emerged following findings in the \textit{graduated dark energy} (gDE) model~\cite{Akarsu:2019hmw}, which demonstrated that a rapid smooth transition from an AdS-like dark energy to a dS-like dark energy at $z \sim 2$ could address the $H_0$ and BAO Ly-$\alpha$ discrepancies~\cite{Akarsu:2019hmw}. It involves replacing the usual cosmological constant ($\Lambda$) of the standard $\Lambda$CDM model with a sign-switching cosmological constant, which can typically be described by sigmoid functions, such as the well-known smooth approximation of the signum function, ${\rm sgn}\,x \approx \tanh{kx}$, for a constant $k > 1$, where $x$ can represent either redshift ($z$) or scale factor ($a=1/(1+z)$, assuming Robertson-Walker metric). For instance, $\Lambda_{\rm s}(z) = \Lambda_{\rm dS} \tanh\qty[\eta(z_{\dagger}-z)]$, where $\eta>1$ determines the rapidity of the transition, and ${\Lambda_{\rm dS}=\Lambda_{\rm s0}/\tanh[\eta\,z_\dagger]}$. For a fast transition (e.g., for $\eta\gtrsim10$) around $z_\dagger\sim1.8$, one can safely take $\Lambda_{\rm dS}\approx\Lambda_{\rm s0}$. 
In the limit as $\eta\rightarrow\infty$, we approach the \textit{abrupt} $\Lambda_{\rm s}$CDM model, which has been commonly investigated in the literature~\cite{Akarsu:2021fol, Akarsu:2022typ, Akarsu:2023mfb}, presenting a one parameter extension of the standard $\Lambda$CDM model; namely,
\begin{equation}
    \textrm{(abrupt)}\,\,\, \textrm{$\Lambda_{\rm s}$CDM:}\,\,\, \Lambda_{\rm s}(z)\rightarrow\Lambda_{\rm s0}\,{\rm sgn}[z_\dagger-z]\,\,\,\textnormal{for}\,\,\, \eta\rightarrow\infty
    \label{eq:ssdeff}
\end{equation}
where $\Lambda_{\rm s0} > 0$ is the present-day value of $\Lambda_{\rm s}(z)$, serving as an idealized depiction of a rapid mirror AdS-dS transition. Originally, this limit case of the model was considered phenomenologically within the framework of general relativity (GR) in~\cite{Akarsu:2021fol, Akarsu:2022typ, Akarsu:2023mfb}. However, without a model, i.e., without an explicit Lagrangian, the paradigm could not be checked against other observables, such as solar system constraints or cosmological linear perturbation theory. Recently, the $\Lambda_{\rm s}$CDM idea was realized within a type II minimally modified gravity model, VCDM. Henceforth, we refer to the original idea based on GR (that is, conjecturing no change in the dynamics of the linear perturbation equations) as $\Lambda_{\rm s}$CDM, and the new realization within the model of VCDM as $\Lambda_{\rm s}$VCDM~\cite{Akarsu:2024qsi}. In this paper, we consider a smooth (implied by finite $\eta$) $\Lambda_{\rm s}$VCDM model that exhibits a \textit{quiescent mirror AdS-dS transition}, to be compared with the standard $\Lambda$CDM and abrupt $\Lambda_{\rm s}$CDM~\eqref{eq:ssdeff} models, using the following functional for $\Lambda_{\rm s}(a)$:
\begin{equation}
\textrm{$\Lambda_{\rm s}$VCDM:}\quad\Lambda_{\rm s}(a)=\Lambda_{\rm dS}\tanh[\zeta(a/a_\dagger-1)]\,,
\label{eq:vcdm1}
\end{equation}
where we fix $\zeta=10^{1.5}$ to study a fast transition that mimics the background of the abrupt $\Lambda_{\rm s}$CDM model as closely as possible while maintaining the same number of free parameters as the abrupt $\Lambda_{\rm s}$CDM, with only one additional parameter, $z_\dagger$, determining the AdS-dS transition redshift, compared to the standard $\Lambda$CDM \footnote{Larger finite values of $\zeta$ are in principle possible but would not be distinguishable with the cosmological data currently available.}\textsuperscript{,}\footnote{Note that, for $\zeta=\eta(1+z)$,~\cref{eq:vcdm1} is equivalent to $\Lambda_{\rm s}(z) = \Lambda_{\rm dS} \tanh\qty[\eta(z_{\dagger}-z)]$, but since both $\eta$ and $\zeta$ are parameters that are relevant around $z\sim z_\dagger$ for a very rapid transition as assumed here, this seemingly dynamic transformation is effectively a simple scaling, $\zeta\approx\eta(1+z_\dagger)$.}. The primary distinction between the $\Lambda_{\rm s}$CDM and $\Lambda_{\rm s}$VCDM models considered here is that $\Lambda_{\rm s}$VCDM is explicitly derived from a well-defined Lagrangian, which uniquely characterizes the model, whereas $\Lambda_{\rm s}$CDM does not possess a Lagrangian formulation. For a detailed theoretical construction of the $\Lambda_{\rm s}$VCDM model, we refer readers to Ref.~\cite{Akarsu:2024qsi}, and for the VCDM theory in which it is embedded, to Refs.~\cite{DeFelice:2020eju, DeFelice:2020cpt}, while a concise overview is provided in~\cref{sec:app1} for convenience.

\section{Datasets and Methodology}
\label{sec:datasets}

To constrain the model parameters, we performed Markov Chain Monte Carlo (MCMC) analyses using a modified version of the publicly available \texttt{CLASS}+\texttt{MontePython} code~\cite{blas2011cosmic, audren2013conservative, Brinckmann:2018cvx}. We employed the ${R-1 < 0.01}$ Gelman-Rubin criterion~\cite{Gelman:1992zz} to ensure the convergence of our MCMC chains. We analyzed the samples using the GetDist Python module.

Our parameter space consists of six parameters common with the standard $\Lambda$CDM model, namely, the present-day physical density parameters of baryons $\omega_{\rm b} \doteq \Omega_{\rm b}h^2$ and cold dark matter (CDM) $\omega_{\rm cdm} \doteq \Omega_{\rm cdm}h^2$, the angular size of the sound horizon at recombination $\theta_{\rm s}$, the amplitude of the primordial scalar perturbation $\log(10^{10}A_{\rm s})$, the scalar spectral index $n_{\rm s}$, and the optical depth $\tau_{\rm reio}$. Additionally, we consider the redshift $z_\dagger$ at which the sign-switching of $\Lambda_{\rm s}$ occurs. We use flat priors for all parameters in our statistical analyses: $\omega_{\rm b} \in [0.018, 0.024]$, $\omega_{\rm cdm} \in [0.10, 0.14]$, $100\,\theta_{\rm s} \in [1.03, 1.05]$, $\ln(10^{10}A_{\rm s}) \in [3.0, 3.18]$, $n_{\rm s} \in [0.9, 1.1]$, $\tau_{\rm reio} \in [0.04, 0.125]$, and $z_\dagger \in [1, 3]$.

The datasets used are as follows:
\begin{itemize}[nosep,wide]
\item CMB: The CMB dataset from the Planck 2018 legacy release is a comprehensive dataset, widely recognized for its precision and accuracy. We use CMB temperature anisotropy and polarization power spectra measurements, their cross-spectra, and lensing power spectrum~\cite{Planck:2019nip,Planck:2018lbu}, namely, the high-$\ell$ \texttt{Plik} likelihood for TT ($30 \leq \ell \leq 2508$) as well as TE and EE ($30 \leq \ell \leq 1996$), the low-$\ell$ TT-only likelihood ($2 \leq \ell \leq 29$) based on the \texttt{Commander} component-separation algorithm in pixel space, the low-$\ell$ EE-only likelihood ($2 \leq \ell \leq 29$) using the \texttt{SimAll} method, and measurements of the CMB lensing. We refer to this dataset as \texttt{Planck}.

\item BAO: We utilize the baryon acoustic oscillation (BAO) measurements reported in~\cref{tab:BAO_measurements}, which consists of both isotropic and anisotropic BAO measurements. The isotropic BAO measurements are identified as $D_{\rm V}(z)/r_{\rm d}$, where $D_{\rm V}(z)$ characterizes the spherically averaged volume distance, and $r_{\rm d}$ represents the sound horizon at the baryon drag epoch. The anisotropic BAO measurements encompass $D_{\rm M}(z)/r_{\rm d}$ and $D_{\rm H}(z)/r_{\rm d}$, where $D_{\rm M}(z)$ denotes the comoving angular diameter distance and $D_{\rm H}(z)$ is expressed as $c/H(z)$, indicating the Hubble distance. These measurements, at eight different effective redshifts, have been derived from the extensive observations conducted by the SDSS collaboration and continuously refined over the past 20 years~\cite{eBOSS:2020yzd}. We refer to this dataset as \texttt{BAO}.
In some analyses, we also consider the partial dataset of 7 BAO measurements with redshift ${z>0.8}$ from~\cref{tab:BAO_measurements}, referred to as \texttt{BAO(${z>0.8}$)}. As demonstrated in~\cite{uzan2023hubble,pitrou2023hubble}, the constraints derived from the individual high-$z$ and low-$z$ BAO datasets yield noticeably different correlations in the $H_0$-$\Omega_{\rm m}$ plane. Given that the correlation in this plane is crucial for elucidating the $H_0$ tension, we opted to limit our BAO sample to redshifts greater than 0.8. Additionally, we included the data point $D_{\rm M}(z)/r_{\rm d}=19.51\pm0.41$ at $z_{\rm eff}=0.85$ obtained with the BAO feature from galaxy clustering in the completed Dark Energy Survey (DES), consisting of six years (Y6) of observations~\cite{DES:2024cme}.

\item SNe Ia: We incorporate the most recent SH0ES Cepheid host distance anchors~\cite{Riess:2021jrx} into the likelihood function by integrating distance modulus measurements of SNe Ia taken from the Pantheon+ sample~\cite{Scolnic:2021amr}. The 1701 light curves in the Pantheon+ dataset correspond to 1550 different SNe Ia events over the redshift range $z \in [0.001, 2.26]$. We refer to this dataset as \texttt{PP\&SH0ES}.

\item Cosmic Shear: We use KiDS-1000 data~\cite{Kuijken:2019gsa,Giblin:2020quj}, which include the weak lensing two-point statistics data for both the auto and cross-correlations across five tomographic redshift bins~\cite{Hildebrandt:2020rno}. We employ the public likelihood in \href{https://github.com/BStoelzner/KiDS-1000_MontePython_likelihood}{KiDS-1000 Montepython likelihood}, and follow the KiDS team analysis, adopting the COSEBIs (Complete Orthogonal Sets of E/B-Integrals) likelihood in our analysis~\cite{KiDS:2020suj}. For the prediction of the matter power spectrum, we use the augmented halo model code, HMcode~\cite{Mead:2015yca}. We highlight that at the level of linear perturbation theory and Boltzmann equations, $\Lambda_{\rm s}$CDM has the same shape as predicted by $\Lambda$CDM. The only effect on the matter power spectrum comes from the $H(z)$ behavior at late times.
As HMcode is robustly tested at the percent level for variations in $H(z)$ functions beyond $\Lambda$CDM, we conclude that no further change to HMcode is necessary to apply cosmic shear measurements on $\Lambda_{\rm s}$CDM. On the other hand, for $\Lambda_{\rm s}$VCDM, we highlight that at the level of linear perturbation theory and Boltzmann equations, the model described here is well-modeled, in the sense that the theory possesses a Lagrangian leading to unique evolution once the background evolution is given. Furthermore, as discussed in~\cite{Akarsu:2024qsi}, $\Lambda_{\rm s}$VCDM does not change the behavior of the spatial components of the Einstein equations, i.e., the lensing equation, for the two gravitational potentials. $\Lambda_{\rm s}$VCDM will affect the time-time component of the field equations, but we expect these effects to be prevalent only on large scales. Since the theory is minimal by construction, the auxiliary field does not propagate, preventing it from becoming unstable. In the small-scale regime, all no-ghost conditions for both matter fields are trivially satisfied.
Therefore, we can use the same nonlinear scale results found in GR and its minimal variation as implemented by default in the HMcode model. We refer to this dataset as \texttt{KiDS-1000}.
\end{itemize}

\begin{table}[t!]
\caption{\rm Clustering measurements for each of the BAO samples from SDSS Ref.~\cite{eBOSS:2020yzd} and DES Y6~\cite{DES:2024cme}.
}
\centering
\scalebox{0.93}{
\begin{tabular}{l|c|c|c|c}

\hline
\textbf{Parameter} & $\bm{\,\,z_{\rm eff}\,\,}$  &  $\bm{\,\,D_{\rm V}(z)/r_{\rm d}\,\,}$ & $\bm{\,\,D_{\rm M}(z)/r_{\rm d}\,\,}$ & $\bm{\,\,D_{\rm H}(z)/r_{\rm d}\,\,}$  \\
\hline

\hline
MGS & 0.15 & $4.47 \pm 0.17$ & --- & --- \\

BOSS Galaxy & $0.38$ & --- & $10.23 \pm 0.17$ & $25.00 \pm 0.76$ \\

BOSS Galaxy & $0.51$ & --- & $13.36 \pm 0.21$ & $22.33 \pm 0.58$ \\

eBOSS LRG & $0.70$ & --- & $17.86 \pm 0.33$ & $19.33 \pm 0.53$ \\

eBOSS ELG & $0.85$ & $18.33_{-0.62}^{+0.57}$ & --- & --- \\

DES Y6 BAO & $0.85$ & --- & $19.51\pm0.41$ & --- \\

eBOSS Quasar & $1.48$ & --- & $30.69 \pm 0.80$ & $13.26 \pm 0.55$ \\

Ly$\alpha$-Ly$\alpha$ & $2.33$ & --- & $37.6 \pm 1.9$ & $8.93 \pm 0.28$ \\

Ly$\alpha$-Quasar & $2.33$ & --- &$37.3 \pm 1.7$ & $9.08 \pm 0.34$ \\
\hline
\hline
\end{tabular}
}
\label{tab:BAO_measurements}
\end{table}

\section{Results and discussion} \label{sec:results}

\begin{table*}[hpt!]
     \caption{Marginalized constraints (mean values with 68\% CL limits) for the free and selected derived parameters of the $\Lambda_{\rm s}$CDM, $\Lambda_{\rm s}$VCDM, and  $\Lambda$CDM models across different dataset combinations. The relative best fit $\Delta \chi^2_{\text{min}}$, Akaike information criterion $\Delta\text{AIC}$, and log-Bayesian evidence $\Delta\ln \mathcal{Z}$ are also provided; negative values indicate a preference for the $\Lambda_{\rm s}$CDM/$\Lambda_{\rm s}$VCDM models over the standard $\Lambda$CDM model.}
     \label{tab:results}
     \scalebox{0.80}{
 \begin{tabular}{lcccccc}
  	\hline
    \toprule
    \textbf{Dataset } & \textbf{Planck}& \textbf{Planck+BAO(z$>$0.8)}\;\; & \textbf{Planck+BAO} \;\; & \textbf{Planck}\;& \textbf{Planck+BAO(z$>$0.8)}\;\; & \textbf{Planck+BAO}   \\
 &  & &  & \textbf{+PP\&SH0ES}& \textbf{+PP\&SH0ES}\;\; & \textbf{+PP\&SH0ES}   \\  \hline
      \textbf{Model} &
      \textbf{$\bm{\Lambda}_{\textbf{s}}$CDM}
      &\textbf{$\bm{\Lambda}_{\textbf{s}}$CDM}
       &\textbf{$\bm{\Lambda}_{\textbf{s}}$CDM}
       &\textbf{$\bm{\Lambda}_{\textbf{s}}$CDM}
       &\textbf{$\bm{\Lambda}_{\textbf{s}}$CDM}
       &\textbf{$\bm{\Lambda}_{\textbf{s}}$CDM}\\
        & \textcolor{teal}{\textbf{$\bm{\Lambda}_{\textbf{s}}$VCDM}} 
         & \textcolor{teal}{\textbf{$\bm{\Lambda}_{\textbf{s}}$VCDM}} 
        & \textcolor{teal}{\textbf{$\bm{\Lambda}_{\textbf{s}}$VCDM}} 
        & \textcolor{teal}{\textbf{$\bm{\Lambda}_{\textbf{s}}$VCDM}} 
        & \textcolor{teal}{\textbf{$\bm{\Lambda}_{\textbf{s}}$VCDM}} 
        & \textcolor{teal}{\textbf{$\bm{\Lambda}_{\textbf{s}}$VCDM}}\\
        & \textcolor{blue}{\textbf{$\bm{\Lambda}$CDM}} 
         & \textcolor{blue}{\textbf{$\bm{\Lambda}$CDM}} 
        & \textcolor{blue}{\textbf{$\bm{\Lambda}$CDM}} 
        & \textcolor{blue}{\textbf{$\bm{\Lambda}$CDM}} 
        & \textcolor{blue}{\textbf{$\bm{\Lambda}$CDM}} 
        & \textcolor{blue}{\textbf{$\bm{\Lambda}$CDM}} 
          \\ \hline
      \vspace{0.1cm}

{\boldmath$10^{2}\omega_{\rm b}$}&$2.241\pm 0.015$ & $2.243\pm 0.015$& $2.237\pm 0.014 $& $2.246\pm 0.015$  &$2.245^{+0.015}_{-0.013}   $& $2.245^{+0.015}_{-0.013}   $\\
& \textcolor{teal}{ $2.250\pm 0.015$} & \textcolor{teal}{$2.243\pm 0.015 $}& \textcolor{teal}{$2.236\pm 0.014$}& \textcolor{teal}{$2.246\pm 0.015            $}&\textcolor{teal}{$2.249\pm 0.014$}&\textcolor{teal}{$2.244\pm 0.014$}\\
& \textcolor{blue}{ $2.238\pm 0.014$} & \textcolor{blue}{$2.245\pm 0.015 $}& \textcolor{blue}{$2.244\pm 0.014$}&\textcolor{blue}{$2.264\pm 0.014$} &\textcolor{blue}{$2.267\pm 0.014$}&\textcolor{blue}{$2.262\pm 0.013$}\\

\vspace{0.1cm}
{\boldmath$\omega_{\rm cdm}$}&$0.1195\pm0.0012$& $0.1193\pm 0.0011$  & $0.1203\pm 0.0010 $&$0.1190\pm 0.0010          $ & $0.1192\pm 0.0010  $& $0.1197\pm 0.0010 $\\
& \textcolor{teal}{$0.1185\pm 0.0013 $} & \textcolor{teal}{$0.1191^{+0.0012}_{-0.0011}$}& \textcolor{teal}{$0.1203\pm 0.0010$}&\textcolor{teal}{ $0.1190\pm 0.0011          $} &\textcolor{teal}{$0.1191\pm 0.0011$}&\textcolor{teal}{$0.1198\pm 0.0010$} \\
& \textcolor{blue}{$0.1200\pm 0.0012 $} & \textcolor{blue}{$0.1189\pm 0.0011 $}& \textcolor{blue}{$0.1191\pm 0.0009$}& \textcolor{blue}{$0.1173\pm 0.0010$}&\textcolor{blue}{$0.1168\pm 0.0010        $}&\textcolor{blue}{$0.1175\pm 0.0008$} \\

\vspace{0.1cm}
{\boldmath$100\theta_{\rm s}$} &$1.04189\pm 0.00029$& $1.04192\pm 0.00029 $& $1.04185\pm 0.00028 $& $1.04194\pm 0.00029        $&$1.04198\pm 0.00028 $& $1.04195\pm 0.00029$\\
& \textcolor{teal}{$1.04201\pm0.00030$} & \textcolor{teal}{$1.04194\pm 0.00030$}& \textcolor{teal}{$1.04184\pm 0.00029$}&\textcolor{teal}{$1.04197^{+0.00028}_{-0.00032}$}& \textcolor{teal}{$1.04198\pm 0.00029$}&\textcolor{teal}{$1.04190\pm 0.00030$} \\
& \textcolor{blue}{$1.04190^{+0.00027}_{-0.00031} $} & \textcolor{blue}{$1.04198\pm 0.00029$}& \textcolor{blue}{$1.04198\pm 0.00028$}&\textcolor{blue}{$1.04217\pm 0.00028$}& \textcolor{blue}{$1.04223\pm 0.00028        $}&\textcolor{blue}{$1.04216\pm 0.00028$} \\

\vspace{0.1cm}
{\boldmath$\ln(10^{10}A_{\rm s})$}&$3.040\pm 0.014$& $3.041^{+0.011}_{-0.015}   $& $3.037\pm 0.014 $& $3.039^{+0.012}_{-0.014}   $&$3.039^{+0.013}_{-0.015}   $ & $3.040^{+0.012}_{-0.014}   $\\
& \textcolor{teal}{$3.033\pm 0.016$} & \textcolor{teal}{$3.036\pm0.015$}& \textcolor{teal}{$3.034\pm 0.014  $}&\textcolor{teal}{$3.032\pm 0.015            $}& \textcolor{teal}{$3.034\pm 0.015$}&\textcolor{teal}{$3.036\pm0.014$} \\
& \textcolor{blue}{$3.046\pm 0.014$} & \textcolor{blue}{$3.049^{+0.013}_{-0.015}   $}& \textcolor{blue}{$3.048\pm 0.014  $}&\textcolor{blue}{$3.058^{+0.015}_{-0.017}$}& \textcolor{blue}{$3.059\pm 0.015   $}&\textcolor{blue}{$3.056^{+0.014}_{-0.016}   $} \\

\vspace{0.1cm}
{\boldmath$n_{\rm s}         $}&$0.9669\pm 0.0043$& $0.9672\pm 0.0040 $& $0.9645\pm 0.0039 $ & $0.9684\pm 0.0039          $&$0.9676\pm 0.0038 $ & $0.9663\pm 0.0040 $\\
& \textcolor{teal}{$0.9694\pm 0.0044$} & \textcolor{teal}{$0.9678\pm 0.0041 $}& \textcolor{teal}{$0.9646\pm 0.0035$}&\textcolor{teal}{$0.9677\pm 0.0039          $}& \textcolor{teal}{$0.9677\pm0.0038$}&\textcolor{teal}{$0.9660\pm 0.0038$} \\
& \textcolor{blue}{$0.9657\pm 0.0041$} & \textcolor{blue}{$0.9681\pm 0.0039 $}& \textcolor{blue}{$0.9675\pm 0.0036$}&\textcolor{blue}{$0.9722\pm 0.0039$}& \textcolor{blue}{$0.9736^{+0.0036}_{-0.0041}$}&\textcolor{blue}{$0.9720\pm 0.0036$} \\

\vspace{0.1cm}
{\boldmath$\tau_{\rm reio} $}&$0.0528\pm 0.0073$& $0.0536^{+0.0059}_{-0.0078}$& $0.0509\pm 0.0072$&  $0.0530^{+0.0060}_{-0.0074}$&$0.0528^{+0.0055}_{-0.0076}$& $0.0527^{+0.0057}_{-0.0071}$\\
& \textcolor{teal}{$0.0507\pm 0.0076$} &\textcolor{teal}{$0.0514\pm0.0076$}& \textcolor{teal}{$0.0493\pm0.0071$}& \textcolor{teal}{$0.0498^{+0.0078}_{-0.0069}$}&\textcolor{teal}{$0.0504^{+0.0079}_{-0.0070}$}&\textcolor{teal}{$0.0504\pm0.0074$} \\
& \textcolor{blue}{$0.0550\pm 0.0072$} & \textcolor{blue}{$0.0573^{+0.0065}_{-0.0082}$}& \textcolor{blue}{$0.0568^{+0.0066}_{-0.0074}$}& \textcolor{blue}{$0.0630^{+0.0073}_{-0.0087}$}&\textcolor{blue}{$0.0642\pm 0.0079$}&\textcolor{blue}{$0.0620^{+0.0067}_{-0.0084}$} \\

\vspace{0.1cm}
{\boldmath$z_{\dagger}             $}&$> 1.45 $ (95\% CL)& $2.20^{+0.17}_{-0.38}$& $> 2.11 $ (95\% CL)&$1.83^{+0.11}_{-0.19}      $&$1.87^{+0.11}_{-0.18}      $& $2.31^{+0.15}_{-0.36}      $\\
& \textcolor{teal}{ $1.88^{+0.28}_{-0.58}$ [$> 1.20 $ (95\% CL)]} & \textcolor{teal}{$2.12^{+0.23}_{-0.27}$}& \textcolor{teal}{$> 2.06 $ (95\% CL)}&\textcolor{teal}{$1.80^{+0.13}_{-0.18}      $}&\textcolor{teal}{$1.86^{+0.12}_{-0.21}$}&\textcolor{teal}{$2.20^{+0.16}_{-0.23}$} \\
& \textcolor{blue}{-} & \textcolor{blue}{-}& \textcolor{blue}{-}&\textcolor{blue}{-}&\textcolor{blue}{-}&\textcolor{blue}{-} \\

\hline




\vspace{0.1cm}
{\boldmath$H_0 {\rm[km/s/Mpc]}            $}&$70.77^{+0.79}_{-2.70}$& $70.39^{+0.87}_{-1.20} $& $68.92^{+0.46}_{-0.55} $& $72.07\pm 0.88             $&$71.68\pm 0.73 $ & $69.82\pm 0.49 $  \\
& \textcolor{teal}{$73.40^{+1.80}_{-4.60} $} & \textcolor{teal}{$70.72^{+0.87}_{-1.30}$}& \textcolor{teal}{$69.10\pm 0.55 $}&\textcolor{teal}{$72.25\pm 0.91             $}&\textcolor{teal}{$71.86\pm 0.79 $}&\textcolor{teal}{$70.01\pm0.50$} \\
& \textcolor{blue}{$67.39\pm 0.55 $} & \textcolor{blue}{$67.88\pm 0.51 $}& \textcolor{blue}{$67.82\pm 0.41 $}&\textcolor{blue}{$68.69\pm 0.47$}&\textcolor{blue}{$68.91\pm 0.46 $}&\textcolor{blue}{$68.62^{+0.34}_{-0.38}$} \\


\vspace{0.1cm}
{\boldmath$\Omega_{\rm m}  $}&$0.2860^{+0.0230}_{-0.0099}  $& $0.2880^{+0.0100}_{-0.0088}  $ & $0.3018\pm 0.0056$&$0.2738\pm 0.0072          $ &$0.2770\pm 0.0063  $& $0.2929^{+0.0044}_{-0.0051}$ \\
& \textcolor{teal}{$0.2650^{+0.0340}_{-0.0190}$} & \textcolor{teal}{$0.2850^{+0.0110}_{-0.0091}$}& \textcolor{teal}{$0.3001\pm 0.0059 $}&\textcolor{teal}{$0.2725\pm 0.0073          $}&\textcolor{teal}{$0.2755\pm 0.0065$}&\textcolor{teal}{$0.2915\pm 0.0049$} \\
& \textcolor{blue}{$0.3151\pm 0.0075$} & \textcolor{blue}{$0.3083\pm 0.0067 $}& \textcolor{blue}{$0.3091\pm 0.0054 $}&\textcolor{blue}{$0.2981\pm 0.0060$}&\textcolor{blue}{$0.2952\pm 0.0058$}&\textcolor{blue}{$0.2989\pm 0.0046$} \\

\vspace{0.1cm}
{\boldmath$\sigma_8         $}& $0.8210^{+0.0064}_{-0.0110} $& $0.8169^{+0.0062}_{-0.0070}$& $0.8143\pm 0.0062 $& $0.8228\pm 0.0068          $& $0.8215\pm 0.0067 $& $0.8160\pm 0.0065 $\\
& \textcolor{teal}{$0.8620^{+0.0160}_{-0.0380}$} & \textcolor{teal}{$0.8414^{+0.0094}_{-0.0140}$}& \textcolor{teal}{$0.8316\pm 0.0078 $}&\textcolor{teal}{$0.8560\pm 0.0120            $}&\textcolor{teal}{$0.8520\pm 0.0110 $}&\textcolor{teal}{$0.8385\pm0.0090$} \\
& \textcolor{blue}{$0.8121^{+0.0055}_{-0.0061}$} & \textcolor{blue}{$0.8098\pm 0.0061 $}& \textcolor{blue}{$0.8097\pm 0.0058 $}&\textcolor{blue}{$0.8085^{+0.0060}_{-0.0070}$ }&\textcolor{blue}{$0.8077\pm 0.0062 $}&\textcolor{blue}{$0.8085^{+0.0060}_{-0.0067}$} \\

\vspace{0.1cm}
{\boldmath$S_8 $}& $0.801^{+0.026}_{-0.016}   $ & $0.800\pm 0.014 $&$0.817\pm 0.010$& $0.786\pm 0.011            $&$0.789\pm 0.010 $& $0.806^{+0.009}_{-0.010}$\\
& \textcolor{teal}{$0.808^{+0.021}_{-0.017}$} & \textcolor{teal}{$0.819\pm 0.012 $}& \textcolor{teal}{$0.832\pm0.011$}&\textcolor{teal}{$0.815\pm 0.011            $}&\textcolor{teal}{$0.816^{+0.011}_{-0.012} $}&\textcolor{teal}{$0.826\pm 0.011 $} \\
& \textcolor{blue}{$0.832\pm 0.013$} & \textcolor{blue}{$0.821\pm 0.012 $}& \textcolor{blue}{$0.822\pm0.010$}&\textcolor{blue}{$0.806\pm 0.011$  }&\textcolor{blue}{$0.801\pm 0.011 $}&\textcolor{blue}{$0.807\pm 0.010 $} \\

\vspace{0.1cm}
{\boldmath$t_0   {\rm[Gyr]}         $}& $13.62^{+0.12}_{-0.04}$ & $13.64^{+0.05}_{-0.04}  $& $13.70^{+0.03}_{-0.02}  $& $13.56\pm 0.04 $& $13.58\pm 0.03 $ & $13.66\pm 0.03$\\

& \textcolor{teal}{$13.52^{+0.18}_{-0.09}$} & \textcolor{teal}{$13.62^{+0.06}_{-0.04}$}& \textcolor{teal}{$13.69\pm 0.03$}&\textcolor{teal}{$13.56\pm 0.04$} &\textcolor{teal}{$13.57\pm 0.04 $}&\textcolor{teal}{$13.65\pm 0.03$} \\
& \textcolor{blue}{$13.79\pm 0.02$} & \textcolor{blue}{$13.78\pm 0.02 $}& \textcolor{blue}{$13.78\pm 0.02$}&\textcolor{blue}{$13.75\pm 0.02$}&\textcolor{blue}{$13.74\pm 0.02 $}&\textcolor{blue}{$13.75\pm 0.02$} \\



\hline
\vspace{0.1cm}
{\boldmath$\chi^2_{\rm min}$}&$2778.06$&$2785.48$&$2796.16$&$4082.28$&$4086.42$&$4106.24$\\
& \textcolor{teal}{$2777.36$} & \textcolor{teal}{$2782.92$}& \textcolor{teal}{$2793.42$}&\textcolor{teal}{$4079.60$}&\textcolor{teal}{$4086.34$}&\textcolor{teal}{$4106.30$} \\
& \textcolor{blue}{$2780.52$} & \textcolor{blue}{$2792.14$}& \textcolor{blue}{$2797.44$}&\textcolor{blue}{$4105.80$}&\textcolor{blue}{$4114.24$}&\textcolor{blue}{$4122.20$} \\
\hline
\vspace{0.1cm}
{\boldmath$\Delta\chi^2_{\rm min}$}&$-2.46$ &   $-6.66$  &$-1.28$&$-23.52$&$-27.82$&$-15.96$\\
& \textcolor{teal}{$-3.16$} & \textcolor{teal}{$-9.22$}& \textcolor{teal}{$-4.02$}&\textcolor{teal}{$-26.20$}&\textcolor{teal}{$-27.90$}&\textcolor{teal}{ $-15.90$} \\
\hline
{\boldmath$\Delta{\rm AIC}$} &$-0.46$&$-4.66$&$0.72$&$-21.52$& $-25.82$ & $-13.96$\\
& \textcolor{teal}{$-1.16$} & \textcolor{teal}{$-7.22$}& \textcolor{teal}{$-2.02$}&\textcolor{teal}{$-24.20$}&\textcolor{teal}{ $-25.90$} &\textcolor{teal}{ $-13.90$} \\
\hline
\vspace{0.1cm}
{\boldmath$\ln \mathcal{Z}$} &$-1423.17$&$-1427.41$&$-1432.97$&$-2076.65$&$-2079.93$&$-2089.64$\\
& \textcolor{teal}{$-1422.21$} & \textcolor{teal}{$-1425.95$}& \textcolor{teal}{$-1432.39$}& \textcolor{teal}{$-2074.32$}&\textcolor{teal}{$-2078.44$}&\textcolor{teal}{ $-2088.24$} \\
& \textcolor{blue}{$-1424.45$} & \textcolor{blue}{$-1429.64$}& \textcolor{blue}{$-1433.52$}&\textcolor{blue}{$-2088.18$}&\textcolor{blue}{$-2092.07$}&\textcolor{blue}{ $-2096.01$} \\
\hline
\vspace{0.1cm}
{\boldmath$ \Delta \ln \mathcal{Z}$} & $-1.28$ & $-2.23$ & $-0.55$ & $-11.53$ & $-12.14$ & $-6.37$\\
&\textcolor{teal}{$-2.24$}&\textcolor{teal}{$-3.69$}&\textcolor{teal}{$-1.13$}&\textcolor{teal}{$-13.86$}&\textcolor{teal}{$-13.63$}&\textcolor{teal}{$-7.77$}\\

 \hline
 \hline
\end{tabular}
}
\end{table*}

\begin{figure*}[ht!]
   \includegraphics[width=16cm]{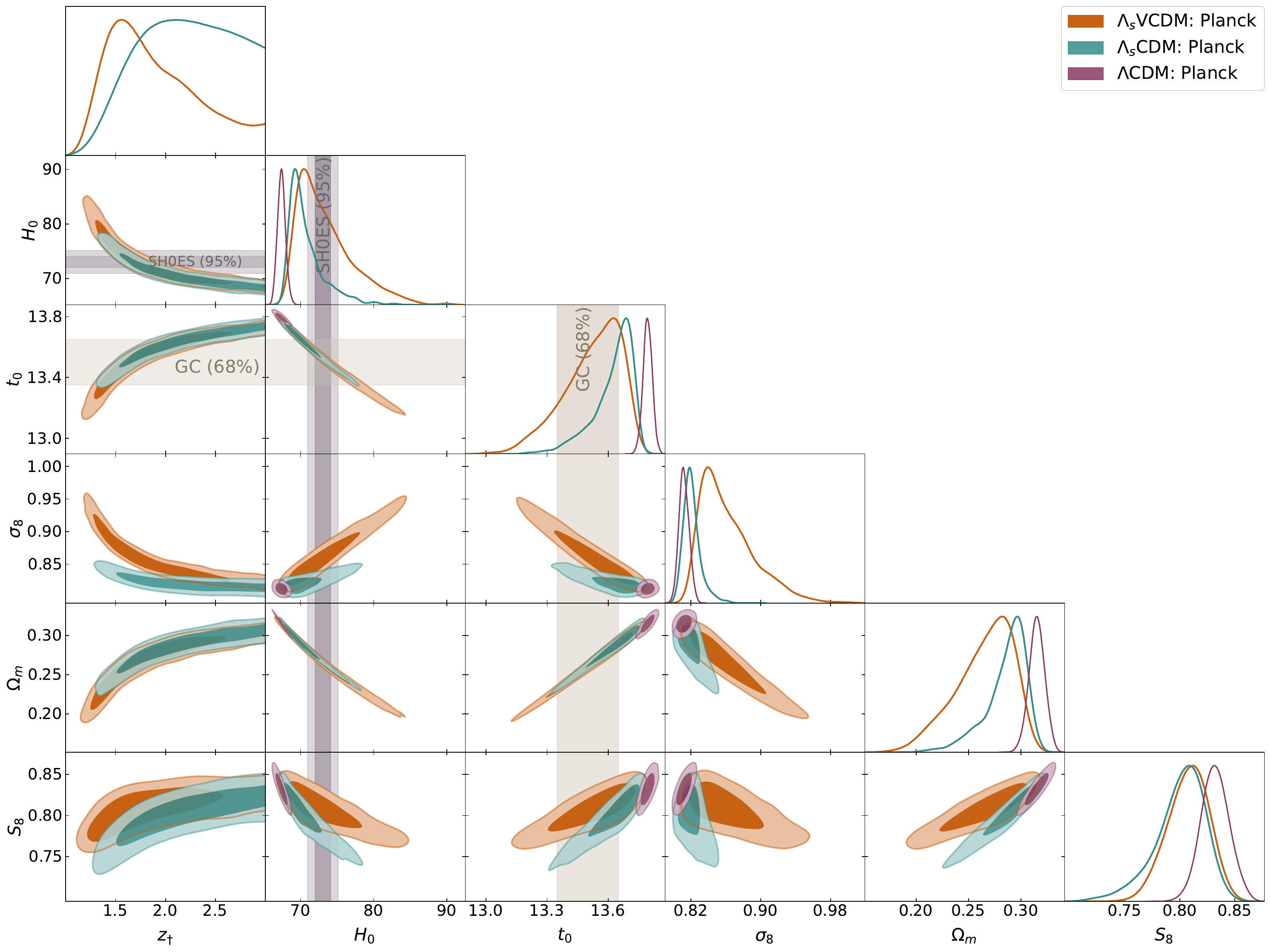}     
    \caption{ One- and two-dimensional (68\% and 95\% CL) marginalized distributions of the $\Lambda_{\rm s}$VCDM, $\Lambda_{\rm s}$CDM, and $\Lambda$CDM model parameters from Planck. The vertical violet and brown bands show the local measurements of $H_0=73.04\pm1.04~{\rm km\, s^{-1}\, Mpc^{-1}}$ (SH0ES)~\cite{Riess:2021jrx} and $t_{0}=13.50\pm0.15\,\rm Gyr$ (stat.)~\cite{Valcin:2021jcg}.
    }
    \label{fig:planck}
\end{figure*}

\begin{figure*}[ht!]
    \centering
  \includegraphics[width=16cm]{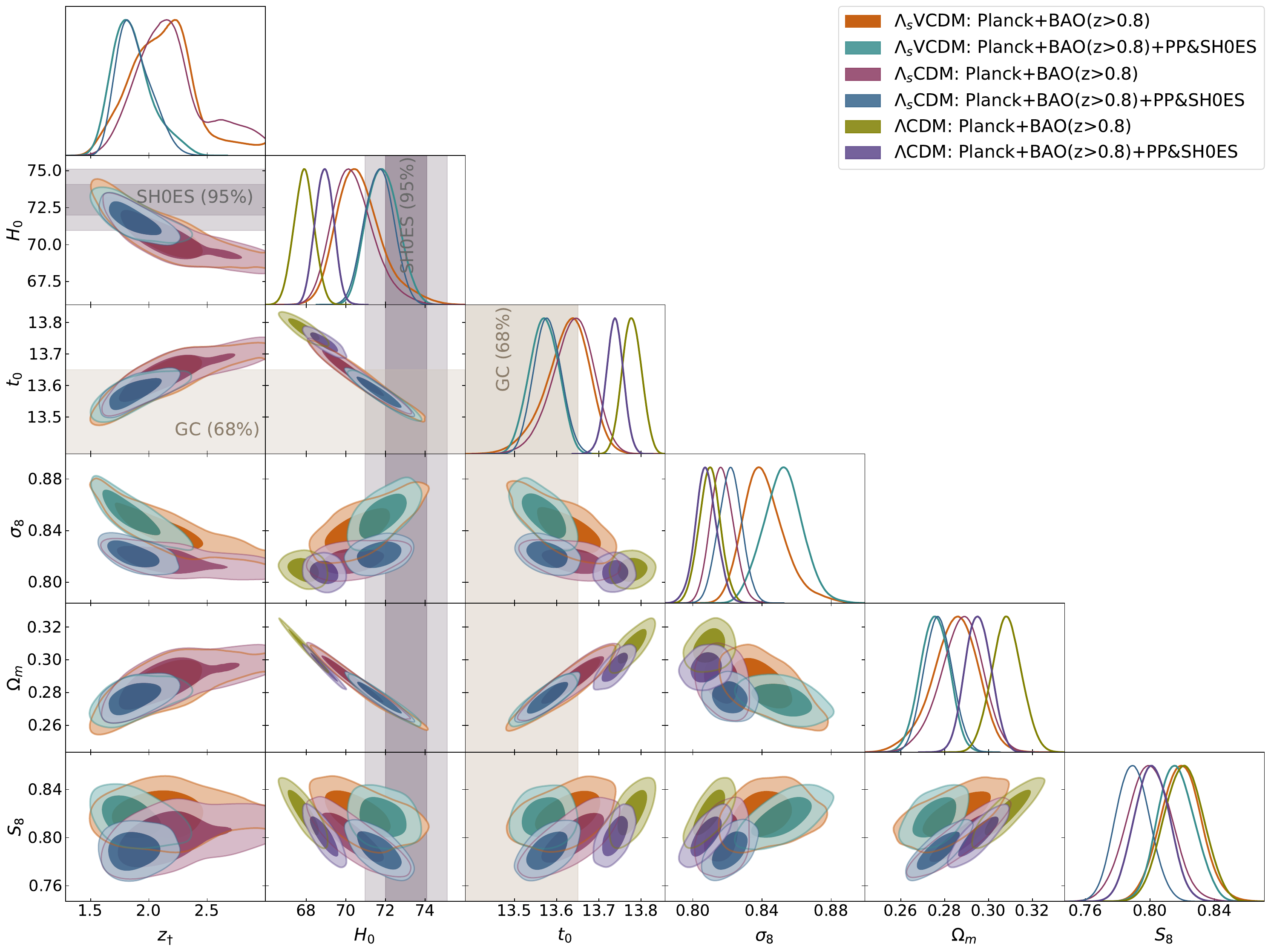}
   \caption{One- and two-dimensional (68\% and 95\% CL) marginalized distributions of the $\Lambda_{\rm s}$VCDM, $\Lambda_{\rm s}$CDM, and $\Lambda$CDM model parameters from Planck+BAO(z$>$0.8) and Planck+BAO(z$>$0.8)+PP$\&$SH0ES. The vertical violet and brown bands show the local measurements of $H_0=73.04\pm1.04~{\rm km\, s^{-1}\, Mpc^{-1}}$ (SH0ES)~\cite{Riess:2021jrx} and $t_{0}=13.50\pm0.15\,\rm Gyr$ (stat.)~\cite{Valcin:2021jcg}.}
   \label{fig:bao1}
\end{figure*}

\begin{figure*}[ht!]
    \centering
  \includegraphics[width=16cm]{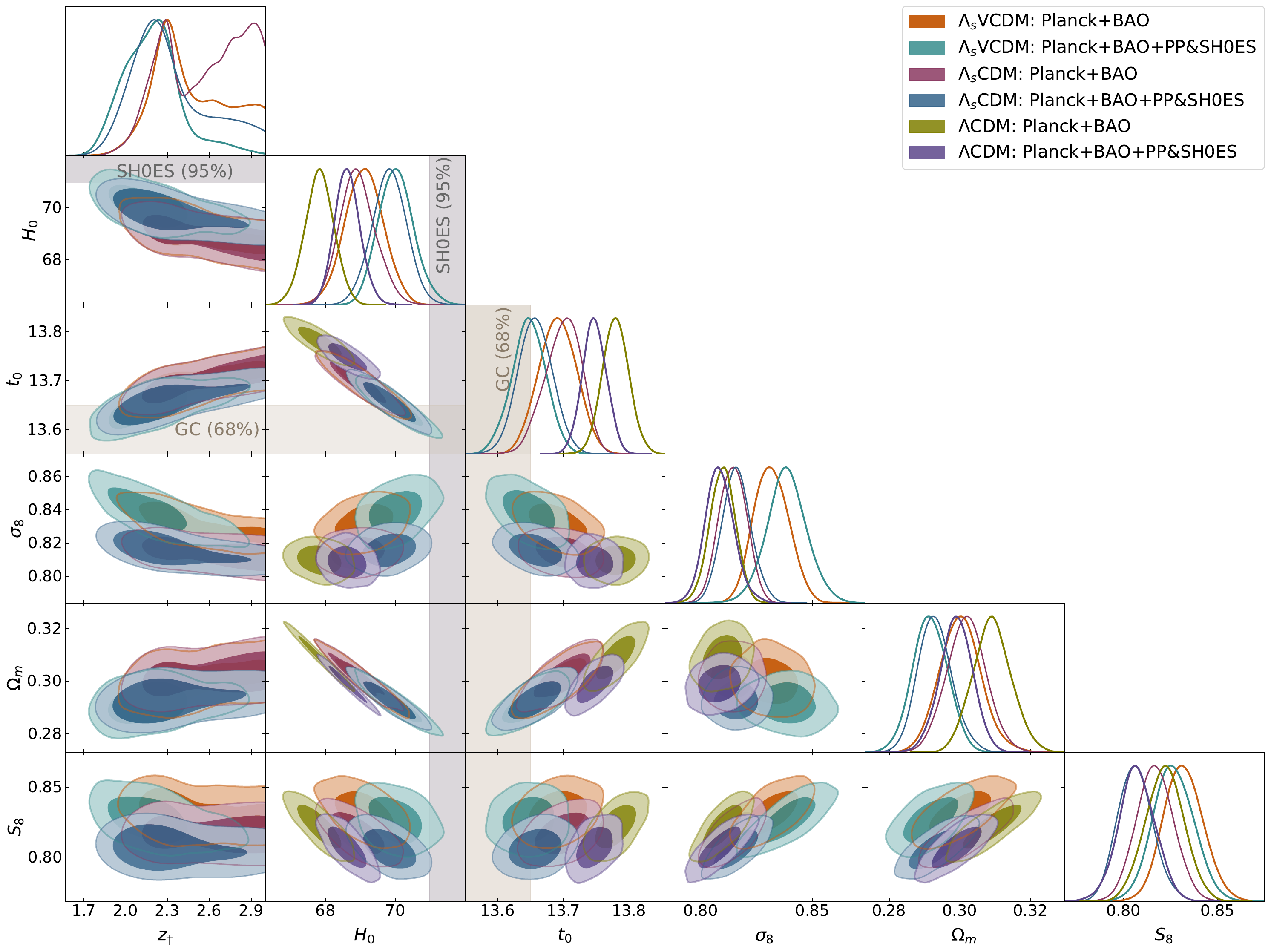}
   \caption{One- and two-dimensional (68\% and 95\% CL) marginalized distributions of the $\Lambda_{\rm s}$VCDM, $\Lambda_{\rm s}$CDM, and $\Lambda$CDM model parameters from Planck+BAO and Planck+BAO+PP$\&$SH0ES. The vertical violet and brown bands show the local measurements of $H_0=73.04\pm1.04~{\rm km\, s^{-1}\, Mpc^{-1}}$ (SH0ES)~\cite{Riess:2021jrx} and $t_{0}=13.50\pm0.15\,\rm Gyr$ (stat.)~\cite{Valcin:2021jcg}.}
   \label{fig:bao2}
\end{figure*}

The marginalized 68\% CL constraints on the baseline free parameters and selected derived parameters for the $\Lambda_{\rm s}$CDM, $\Lambda_{\rm s}$VCDM, and $\Lambda$CDM models are presented in\cref{tab:results}. The Planck, BAO, and PP\&SH0ES datasets are utilized in different combinations of interest. \cref{fig:planck,fig:bao1,fig:bao2} show the one- and two-dimensional marginalized distributions for a few  parameters of interest of the models considered in this work, derived from Planck only dataset, and  its combinations with PP\&SH0ES, BAO (${z>0.8}$), and the full BAO dataset.

As a primary feature of our observational tests, it is notable that the six parameters of the common baseline remain unchanged across all models, with the maximum shift of $\sim 1\sigma$ between $\Lambda_{\rm s}$VCDM and $\Lambda$CDM. When interpreting in terms of derived quantities from these core parameters, it becomes evident that since both scenarios ($\Lambda_{\rm s}$CDM and $\Lambda_{\rm s}$VCDM) can produce a high value for $H_0$, they simultaneously will project a lower value for $\Omega_{\rm m}$ compared to the $\Lambda$CDM model. This is because the CMB accurately measures $\Omega_{\rm m} h^2$ from the peak structure of the damping tail, resulting into a negative correlation in the $H_0$-$\Omega_{\rm m}$ plane. This trend is observed in all analyses carried out, but the effect is more evident in the analysis with CMB only. Previous studies~\cite{Akarsu:2021fol, Akarsu:2022typ,Akarsu:2023mfb,Yadav:2024duq} (see also~\cite{Adil:2023exv}) have shown that these models provide a compelling alternative solution to the $H_0$ tension.

The effects on CMB anisotropies are anticipated to be observed in the amplitude of the late integrated Sachs-Wolfe effect (ISW), particularly manifested at large angular scales.  This effect hinges on the duration of the dark energy-dominated stage, determined by the time of equality between matter and dark energy density, fixed by the ratio $\Omega_{\Lambda_{\rm s}}/\Omega_{\rm m}$, where $\Omega_{\rm m} = \Omega_{\rm b} + \Omega_{\rm cdm}$. A higher $\Omega_{\Lambda_{\rm s}}$ suggests an extended dark energy domination period, consequently amplifying the late integrated Sachs-Wolfe effect. In our primary baseline, constraints on the baryon density are expected to remain practically unchanged. Additionally, in the models under consideration, we assume spatial flatness for our Universe. Thus, at late times, $\Omega_{\Lambda_{\rm s}} = 1 - \Omega_{\rm m}$ (neglecting radiation), and the alterations induced by different constraints on $\Omega_{\rm cdm}$ will primarily govern corrections to CMB anisotropies at large scales. On the other hand, $\Omega_{\rm cdm}$ will influence the amplitude of the third peak in the CMB power spectra and also impact constraints on $H_0$ through the relationship $h \simeq \sqrt{\omega_{\rm m}/(1 - \Omega_{\Lambda_{\rm s}})}$ (assuming $\Omega_{\rm k} = 0$) at late times. Furthermore, alterations in the late-time expansion of the Universe induced by the mirror AdS-dS transition will modify the angular diameter distance at decoupling. The magnitude of these correlations in $H_0$ is directly proportional to the potential values for the mirror AdS-dS transition (see Fig.~2 and explanation in~\cite{Akarsu:2021fol}). Due to the significant degeneracy and correlation between  $z_{\dagger}$ and $H_0$, we conclude that the CMB data alone has limited constraining power for directly determining the transition redshift $z_{\dagger}$. Therefore, it is imperative to complement CMB data with geometric measurements to achieve a comprehensive understanding of $z_{\dagger}$.

Furthermore, we can interpret these results in a physical context, stemming from a fundamental property of the model: a transition from anti--de Sitter to de Sitter dynamics at a redshift of $z_{\dagger} \sim 2$. For the $\Lambda_{\rm s}$VCDM model, following the AdS-dS transition, the introduction of a new auxiliary scalar field results in an effective cosmological constant, leading to a prediction that its density parameter surpasses that of the $\Lambda$CDM model, i.e., $\Omega_{\Lambda_{\rm s}} > \Omega_{\Lambda}$ after transition. Consequently, assuming that the density evolution of other species such as baryons and radiation remains unaffected, this also implies that the cold dark matter density will dilute at a faster rate than expected, thereby predicting lower values for $\Omega_{\rm m}$ as summarized in~\cref{tab:results}.  Then, as the transition indicates a greater influence of effective dark energy and a decrease in cold dark matter density at late times, the expected consequence is that the Hubble parameter, $H(z)$, will be greater than predicted in the $\Lambda$CDM model. In other words, the Universe expands faster in the $\Lambda_{\rm s}$VCDM model after the transition than in the $\Lambda$CDM model. The interpretation of parameters and dynamics for the $\Lambda_{\rm s}$CDM model is the same; however, the transition occurs instantaneously in this scenario. It is worth noting that both models predict a transition redshift $z_{\dagger}$ consistent with each other in all analyses conducted.

\vspace{10pt}

It is interesting to notice the difference in the results for $H_0$ between $\Lambda_{\rm s}$CDM and $\Lambda_{\rm s}$VCDM in the constraints derived solely from Planck data. At the background level, they differ slightly since the transition is smooth, although rapid, for the $\Lambda_{\rm s}$VCDM model. However, at the level of perturbations, the Bardeen potential $\Phi$ changes because its dynamics depends on $a\Lambda_{{\rm s}, a}$, which can assume large values.\footnote{\label{foonote3}The equations of motion governing the dynamics for $\Phi$ can be written as $\Phi_{,\tau}+aH\Psi=\frac32 (a^2/k^2)\,\Gamma\,\sum_I(\varrho_I+p_I)\theta_I$, where $\tau$ is the conformal time, $\Psi$ is the second Bardeen potential, and $\theta_I$ corresponds to the scalar component of the $I$th fluid 3-velocity, $I$ running over all the standard matter fields. Here $\Gamma\equiv [k^2/a^2-3H_{,\tau}/a]/[k^2/a^2+(9/2)\sum_K(\varrho_K+p_K)]$. Since we can rewrite $\Gamma$ as $\Gamma\equiv [k^2/a^2+(9/2)\sum_K(\varrho_K+p_K)-(a/2)\Lambda_{{\rm s}, a}]/[k^2/a^2+(9/2)\sum_K(\varrho_K+p_K)]$, we can deduce a suppression of the numerator at transition (or even a switch of its sign for a very sharp transition). In $\Lambda_{\rm s}$CDM, having set the perturbation dynamics identical to GR, $\Gamma$ is unity.} This will also affect the dynamics of $\delta_{\rm m}$, which explicitly depends on $\dot{\Phi}$. This combination of changes leads to differences between the $\Lambda_{\rm s}$CDM and $\Lambda_{\rm s}$VCDM models, allowing the latter to have larger contours and a mean value for $H_0$ closer to the SH0ES measurement of $H_0=73.04\pm1.04~{\rm km\, s^{-1}\, Mpc^{-1}}$~\cite{Riess:2021jrx}.\footnote{To disentangle these two contributions---the one of the background from the one of the modified perturbation equations---it could be interesting to modify the background of $\Lambda_{\rm s}$CDM exactly match that of $\Lambda_{\rm s}$VCDM, smoothing out the instantaneous transition into a smooth but rapid one. Then we could see how much the smoothing of the background improves the fit of $\Lambda_{\rm s}$CDM, and vice versa, how much the modified perturbation dynamics of $\Lambda_{\rm s}$VCDM affect the results. We leave the discussion of this point for a separate project.} Another way to understand this larger value for $H_0$ is to realize that $z_\dagger$ tends to be smaller in the $\Lambda_{\rm s}$VCDM model compared to the $\Lambda_{\rm s}$CDM model. Notably, for $\Lambda_{\rm s}$VCDM in the Planck-alone case, the lower bound on $z_\dagger$ reaches values of $1.2$ at a 95\% CL (compared to $1.45$ in the $\Lambda_{\rm s}$CDM model), leading to large values of $H_0$ (and thereby, smaller values of $\Omega_{\rm m}$). When the models become indistinguishable from the $\Lambda$CDM model at the high end of $z_\dagger$, they predict $H_0$ values similar to those of the $\Lambda$CDM. This picture places $H_0$ in the range of $75.2~{\rm km\, s^{-1}\, Mpc^{-1}}$ to $68.8~{\rm km\, s^{-1}\, Mpc^{-1}}$, with a mean value of $73.4~{\rm km\, s^{-1}\, Mpc^{-1}}$. This constraint aligns perfectly with the SH0ES measurement of $73.30 \pm 1.04~{\rm km\, s^{-1}\, Mpc^{-1}}$ (derived by including high-$z$ SN Ia)~\cite{Riess:2021jrx}, with the mean value almost exactly matching it, and with the latest measurements of $73.17 \pm 0.86~{\rm km\, s^{-1}\, Mpc^{-1}}$~\cite{Breuval:2024lsv} and $73.22 \pm 0.68~(\text{stat}) \pm 1.28~(\text{sys})~{\rm km\, s^{-1}\, Mpc^{-1}}$~\cite{Uddin:2023iob}. Notably, $H_0$ predictions from the Planck-alone analysis of both $\Lambda_{\rm s}$CDM and $\Lambda_{\rm s}$VCDM models exhibit no tension at all with any of the SH0ES measurements. Specifically, the discrepancy for Planck-$\Lambda_{\rm s}$CDM is only $1-1.2\sigma$, and for Planck-$\Lambda_{\rm s}$VCDM, it is almost nonexistent, at an amazingly low $0.0-0.1\sigma$.

Next, we examine the ramifications of our choices within the BAO sample when considering both $\Lambda_{\rm s}$CDM and $\Lambda_{\rm s}$VCDM. In the joint analysis involving Planck+BAO (${z>0.8}$, including DES Y6), a clear trend emerges toward lower values of $\Omega_{\rm m}\sim0.29$, consequently leading to higher values of $H_0\sim70.4-70.7~{\rm km\, s^{-1}\, Mpc^{-1}}$. This trend is attributed to the distinctive correlation exhibited by the BAO sample within this redshift range~\cite{eBOSS:2020yzd}. In this context, the cosmological framework's capability to anticipate these correlations naturally manifests in this joint analysis. Expanding our scope to encompass all BAO samples, i.e., Planck+BAO, the comprehensive sample unsurprisingly aligns the BAO constraints more closely with the values prescribed by the $\Lambda$CDM model. However, it is noteworthy that both models still predict $H_0\sim69~{\rm km\, s^{-1}\, Mpc^{-1}}$, values slightly higher than those predicted by the $\Lambda$CDM model, thereby reducing the $H_0$ tension. Consequently, our analysis solely with Planck+BAO helps in mitigating the $H_0$ tension. From these joint analyses, evaluating the tension individually using the standard 1D tension metric, we find a tension of $\sim1.8\sigma$ for both models from Planck+BAO(${z>0.8}$) and a tension of $\sim3.3\sigma$ from Planck+BAO. While for the combination Planck+BAO(${z>0.8}$), we can constrain $z_{\dagger}\sim2.1-2.2$ at 68\% CL, when we consider the full BAO data, we find only a lower limit that gives $z_{\dagger}>2.1$ at 95\% CL for both $\Lambda_{\rm s}$CDM and $\Lambda_{\rm s}$VCDM.\footnote{Recent BAO measurements from the DESI collaboration~\cite{DESI:2024uvr, DESI:2024lzq, DESI:2024mwx} have become available, offering potentially new insights into the nature of dark energy. While the completed SDSS-BAO dataset (BOSS+eBOSS)\cite{eBOSS:2020yzd} used in this study has a constraining power comparable to the DESI-BAO dataset\cite{DESI:2024mwx}, the DESI data may provide fresh evidence regarding the dynamical aspects of dark energy (see, e.g.,~\cite{Giare:2024smz, Mukherjee:2024ryz, Ye:2024ywg, Giare:2024gpk, Dinda:2024ktd, Jiang:2024xnu}). However, the models considered in this paper do not account for dynamical dark energy in their background evolution. The sole free parameter in these theories, the transition epoch $z_\dagger$, is already well-constrained by the complete SDSS-BAO dataset. Consequently, incorporating the DESI data is not expected to significantly alter the precision of the baseline parameters or their correlations relative to the SDSS sample. Therefore, our conclusions remain robust even when DESI data are considered. In a future study, we will update the results using the BAO-DESI dataset for performance evaluation purposes. However, the main results and conclusions are anticipated to remain consistent with those presented here.}

\vspace{10pt}

We further investigated the combination of the PP\&SH0ES and Planck datasets. The findings are summarized in~\cref{tab:results}. Consistent with previous analyses, both $\Lambda_{\rm s}$CDM and $\Lambda_{\rm s}$VCDM models yield observational constraints that align well with each other in this combined assessment. Employing the standard 1D tension metric to assess individual tensions, we observe a tension of approximately $0.6\sigma$ on $H_0$, with $H_0\sim72~{\rm km\, s^{-1}\, Mpc^{-1}}$. Consequently, based on this joint analysis, we conclude that both models effectively alleviate the tension in $H_0$ with significant statistical support. It is noteworthy that, as it is well known, the Planck data and PP\&SH0ES data are in tension within the $\Lambda$CDM model. Therefore, combining these two datasets for the $\Lambda$CDM analyses is not statistically worthwhile, but these results are given here for completeness. The same argument applies to Planck+PP\&SH0ES baseline and Planck+BAO; consequently, combining all these datasets is not statistically sound in this scenario. However, with regards to the BAO samples, we can instead only consider the combination of Planck+BAO(${z>0.8}$)+PP\&SH0ES, as these datasets demonstrate internal consistency, i.e., these exhibit no tension among them. The results of our analysis involving Planck+BAO(${z>0.8}$)+PP\&SH0ES are also summarized in~\cref{tab:results}. In this case, we observe only a tension of approximately 1$\sigma$ between the predictions of $\Lambda_{\rm s}$CDM and $\Lambda_{\rm s}$VCDM and the SH0ES $H_0$ measurement. Therefore, based on this comprehensive joint analysis, we can conclude that both models effectively alleviate the $H_0$ tension.

\begin{figure*}[ht!]
    \centering
  \includegraphics[width=5.9cm]{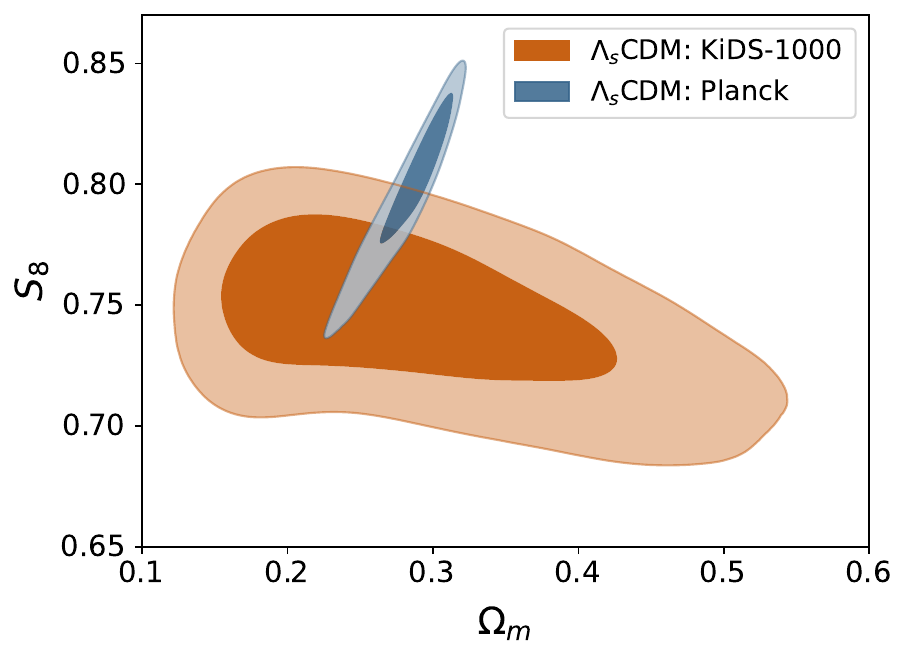}
    \includegraphics[width=5.9cm]{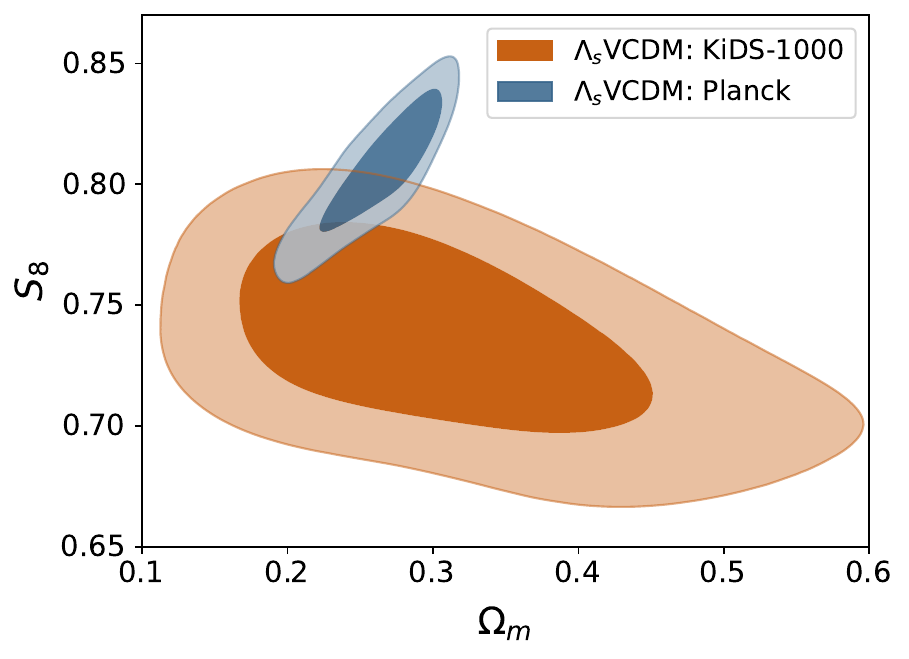} 
    \includegraphics[width=5.9cm]{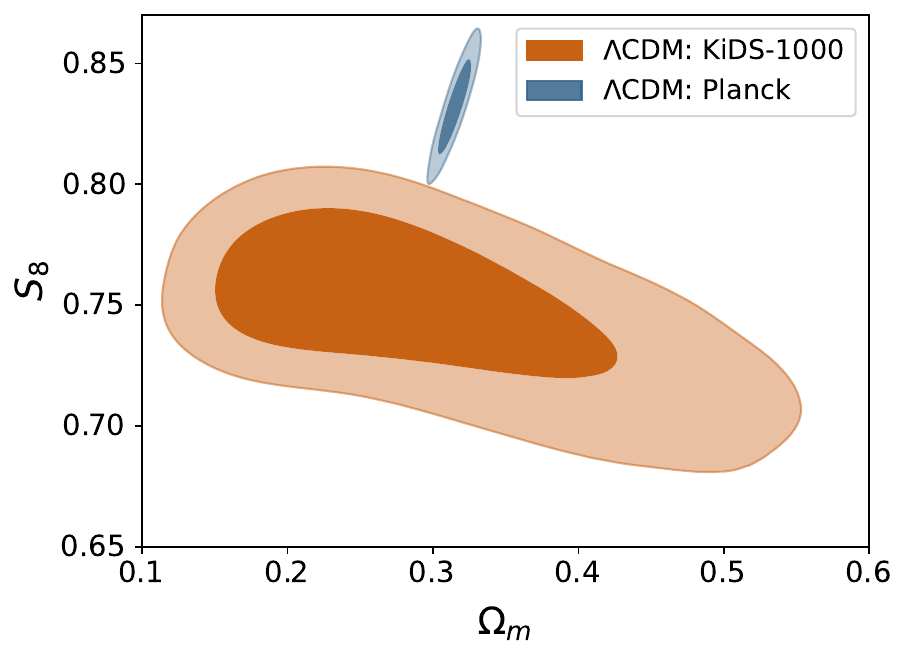} 
    \caption{2D contours (68\% \& 95\% CL) in the $\Omega_{\rm m}$-$S_8$ plane for $\Lambda_{\rm s}$CDM, $\Lambda_{\rm s}$VCDM, and $\Lambda$CDM. ${S_8 = 0.746^{+0.026}_{-0.021}}$ ($\Lambda_{\rm s}$CDM: KiDS-1000), ${S_8 = 0.801^{+0.026}_{-0.016}}$ ($\Lambda_{\rm s}$CDM: Planck), ${S_8 = 0.736\pm 0.027  }$ ($\Lambda_{\rm s}$VCDM: KiDS-1000), $S_8 = 0.808^{+0.021}_{-0.017}$ ($\Lambda_{\rm s}$VCDM: Planck), ${S_8 =0.749^{+0.027}_{-0.020}}$ ($\Lambda$CDM: KiDS-1000), ${S_8 = 0.832\pm 0.013}$ ($\Lambda$CDM: Planck) at 68\% CL. In the $\Lambda_{\rm s}$CDM and $\Lambda_{\rm s}$VCDM models, the $S_8$ tension reduces to significance levels of $1.7\sigma$ and $2.2\sigma$, respectively, compared to a higher level of $3.0\sigma$ in the standard $\Lambda$CDM model.}
    \label{fig:kids}
\end{figure*}

We now turn our attention to the weighted amplitude of matter fluctuations, quantified by the parameter $S_8 = \sigma_8\sqrt{\Omega_{\rm m}/0.3}$ using the standard definition. Initially, it is crucial to acknowledge that in all joint analyses, both the $\Lambda_{\rm s}$CDM and $\Lambda_{\rm s}$VCDM models exhibit a tendency to project higher values for $\sigma_8$ compared to $\Lambda$CDM, with $\Lambda_{\rm s}$VCDM notably predicting a higher value compared to $\Lambda_{\rm s}$CDM. This difference arises from the fact that $\Lambda_{\rm s}$VCDM incorporates linear perturbative effects of scalar modes, whereas the $\Lambda_{\rm s}$CDM model does not. Fundamentally, this distinction arises from the analysis of CMB data, where $\Lambda_{\rm s}$VCDM also impacts the CMB spectrum during late times, i.e., at large angular scales. This influence stems from alterations in the scalar fields $\Phi$ and $\Psi$, their temporal variations, and the background evolution. Consequently, with respected to the standard $\Lambda$CDM model, $\Lambda_{\rm s}$VCDM affects the CMB spectrum in two distinct ways, whereas $\Lambda_{\rm s}$CDM solely influences it through background evolution $H(z)$. Conversely, as discussed previously, it is established that both models predict a lower value for $\Omega_{\rm m}$. In other words, both the $\Lambda_{\rm s}$VCDM and $\Lambda_{\rm s}$CDM models forecast an increased rate of structure formation, yet simultaneously anticipate less matter density parameter today, resulting in overall lower values for $S_8$ compared to $\Lambda$CDM. It is noteworthy that all constraints on $S_8$ are mutually compatible at approximately $\sim$1$\sigma$ CL between $\Lambda_{\rm s}$VCDM and $\Lambda_{\rm s}$CDM.

To correctly assess the $S_8$ tension between the weak lensing measurements and the CMB ones, we conducted an analysis using only KiDS-1000 data~\cite{KiDS:2020suj, Giblin:2020quj} for all the models under consideration because the $S_8$ constraints are model-dependent for this observable. \cref{fig:kids} illustrates the 2D contour plots in the $S_8$-$\Omega_{\rm m}$ plane for the $\Lambda_{\rm s}$CDM, $\Lambda_{\rm s}$VCDM, and $\Lambda$CDM models. In each case, the figure compares the 2D contours between Planck and KiDS-1000 data exclusively for that particular model. As widely recognized, the right panel displays a disagreement in the $S_8$-$\Omega_{\rm m}$ plane for the $\Lambda$CDM model between the two probes, indicating an $S_8$ tension at the $3.0\sigma$ level. Conversely, in the left panel depicting the $\Lambda_{\rm s}$CDM scenario, we observe strong compatibility between the bounds in the $S_8$-$\Omega_{\rm m}$ plane from Planck and KiDS-1000 data considered separately. In this scenario, we derive $S_8 = 0.746^{+0.026}_{-0.021}$ at 68\% CL from the KiDS-1000 sample, while considering the Planck analysis alone, we observe $S_8 = 0.801^{+0.026}_{-0.016}$. Subsequently, we find that the tension between both samples amounts to approximately 1.7$\sigma$. We arrive at similar conclusions for the $\Lambda_{\rm s}$VCDM model, as shown in the middle panel.
We note that both the $\Lambda_{\rm s}$VCDM and $\Lambda_{\rm s}$CDM models do not significantly alter the  KiDS-1000 alone constraints in the $S_8 - \Omega_{\rm m}$ plane compared to $\Lambda$CDM. These models affect the three two-point correlation functions ($3\times2$pt analysis) by primarily altering the background evolution described by $H(z)$, though $\Lambda_{\rm s}$VCDM predicts changes at the linear level of order 1; however, these effects are not significantly distinguishable from those in $\Lambda$CDM. As established in the literature, cosmic shear analyses alone are unable to effectively constrain $H_0$ and $\Omega_{\rm m}$ simultaneously, but they do constrain the parameter $\sigma_8$. Consequently, the constraints in $\Lambda_{\rm s}$CDM are expected to be nearly identical to those in $\Lambda$CDM, as both models are virtually indistinguishable at the level of $\sigma_8$. Therefore, cosmic shear alone will not differentiate the constraints on $H_0$ and $\Omega_{\rm m}$ between these two models. On the other hand, as shown in~\cref{tab:results}, the $\Lambda_{\rm s}$VCDM model, in addition to altering the constraints on $\Omega_{\rm m}$, also predicts changes in $\sigma_8$ due to minimal differences in linear perturbations compared to the $\Lambda$CDM model. In terms of cosmic shear analyses, this introduces an additional degeneracy in $\Omega_{\rm m}$ but keeps the resulting $S_8$ nearly the same as in $\Lambda$CDM. Therefore, we conclude that the extended models considered in this work do not predict significant changes in cosmic shear analyses. Thus, the improvement in the $S_8$ tension in these models essentially arises from modifications in the CMB constraints, which lower the value of $S_8$ to align with predictions from cosmic shear surveys. Specifically, by reducing $\Omega_{\rm m}$, as demonstrated in~\cref{fig:planck}, the positive correlation between the parameters $S_8$ and $\Omega_{\rm m}$ with $z_\dagger$ results in the observed positive correlation between $S_8$ and $\Omega_{\rm m}$ in~\cref{fig:kids} as well.

\begin{figure}[ht!]
    \centering
    \includegraphics[width=8.0cm]{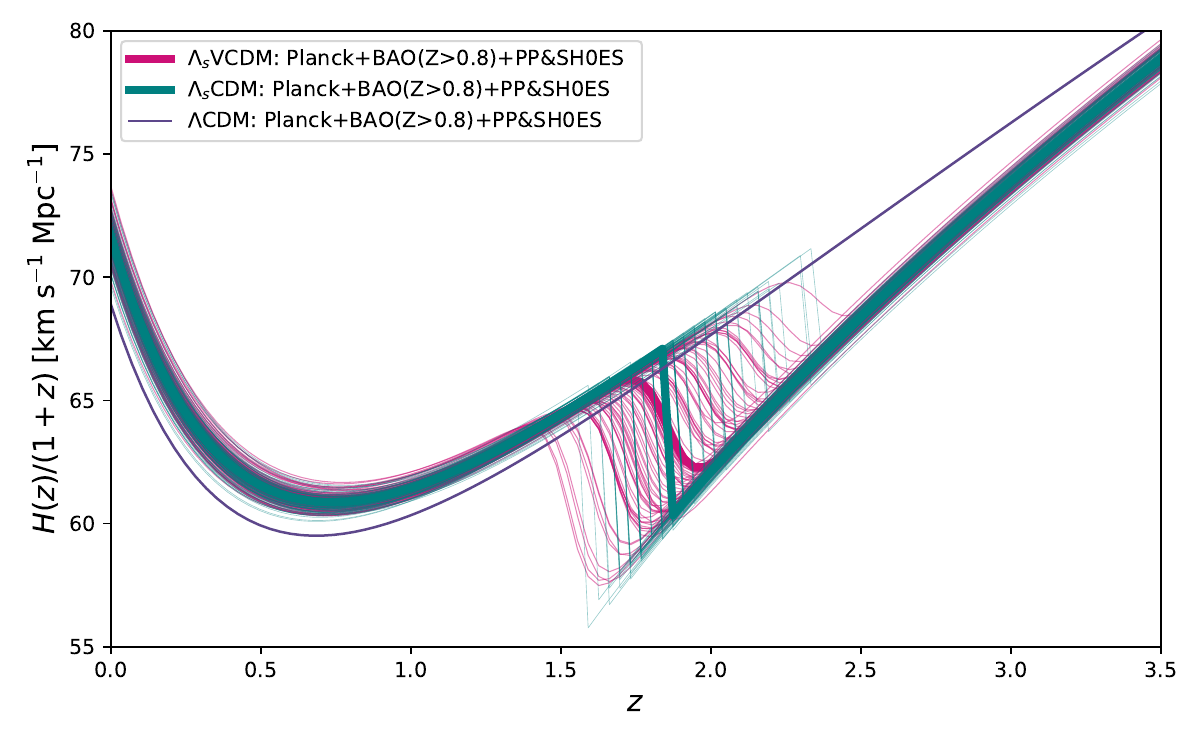}
    \includegraphics[width=8.0cm]{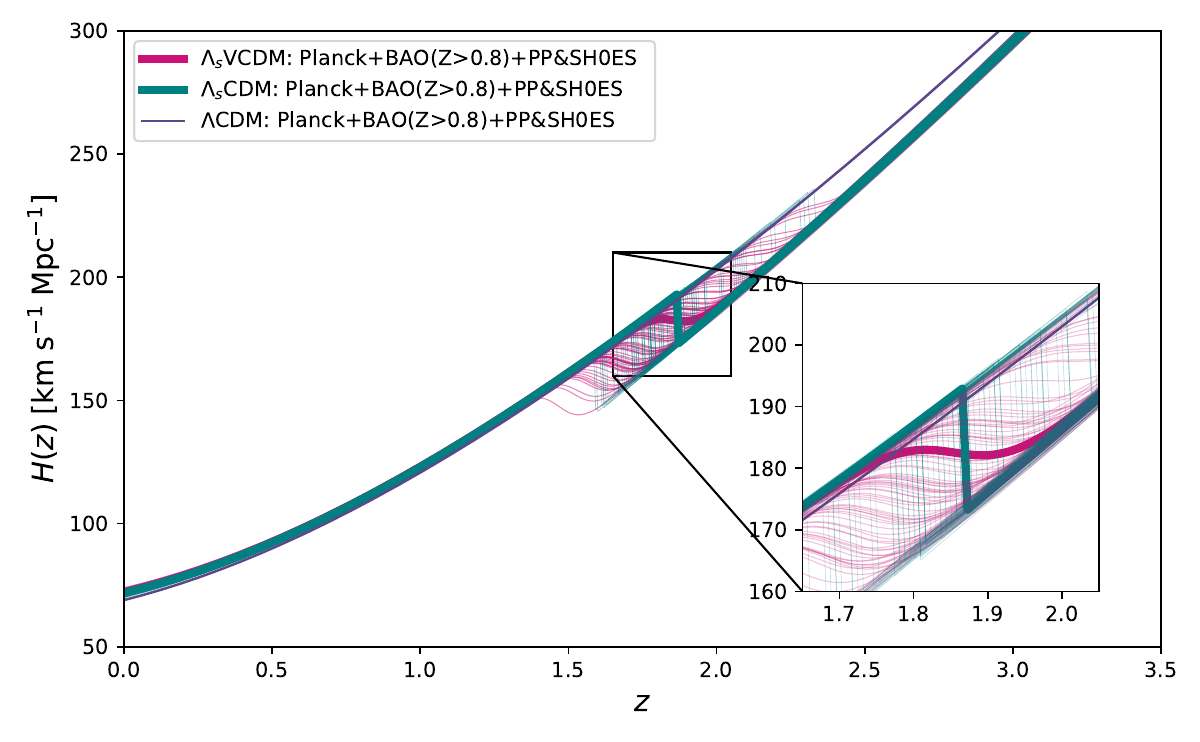}
     \includegraphics[width=9.0cm]{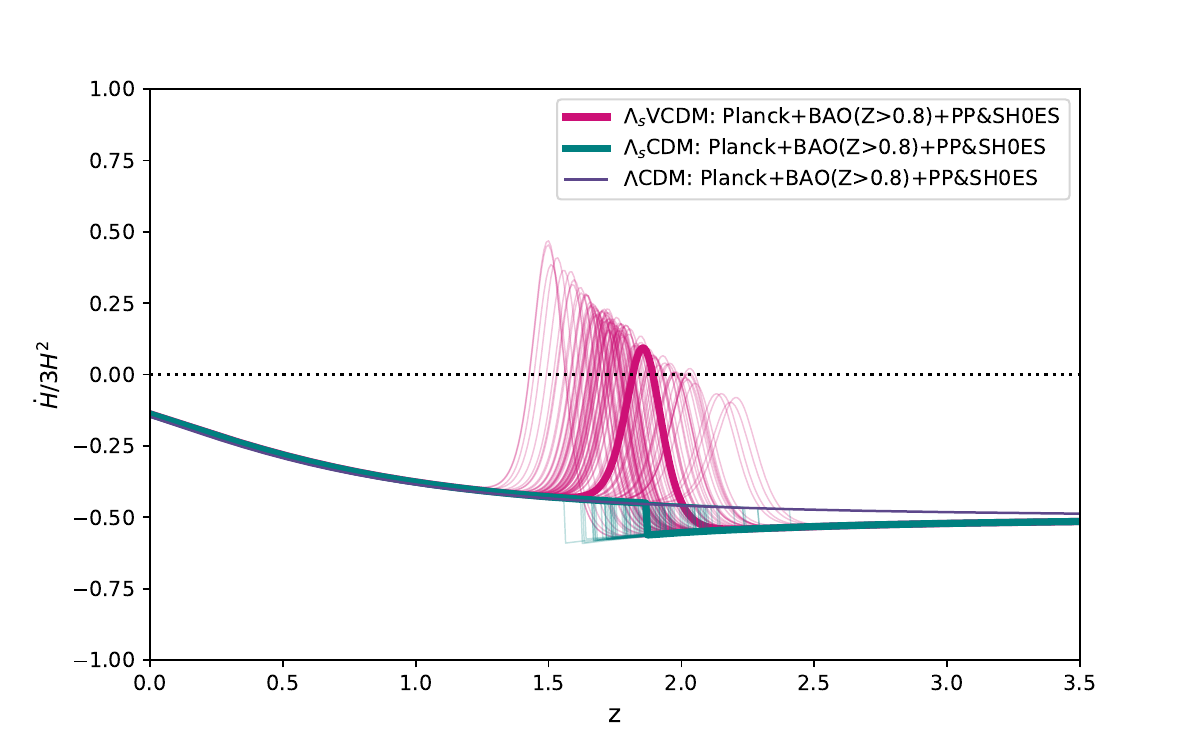}
    \caption{Redshift evolution of the comoving Hubble parameter, $\dot{a} = H(z)/(1+z)$, Hubble parameter, $H(z) = \frac{\dot{a}}{a}$, and the time rate of change of the Hubble parameter scaled by $3H^2$, $\dot{H}/3H^2 = -\frac{1+z}{3H(z)} \, \frac{{\rm d}H(z)}{{\rm d}z}$, for the $\Lambda_{\rm s}$CDM, $\Lambda_{\rm s}$VCDM, and $\Lambda$CDM models. Plotted for the combined Planck+BAO(${z>0.8}$)+PP\&SH0ES dataset.}
    \label{fig:Hz}
\end{figure}

The (present-day) age of the universe measured using the oldest globular clusters (GCs), in a model-agnostic way, suggests $t_{\rm u} = 13.50 \pm 0.15 \, (\text{stat.}) \pm 0.23 \, (\text{sys.})$ Gyr ($\pm 0.27$ with combined uncertainties). This aligns well with the age predicted by the standard $\Lambda$CDM model, though current systematic uncertainties in $t_{\rm u}$ are substantial. Efforts are underway to reduce these uncertainties to better discriminate among cosmological models through the age they predict, especially those addressing the $H_0$ tension~\cite{Vagnozzi:2021tjv}. Considering only statistical uncertainties, the GC-estimated age shows a $\sim2\sigma$ tension with $\Lambda$CDM in all our analyses summarized in~\cref{tab:results}. This level of discrepancy in the age may not indicate a serious issue for many, particularly as long as a model predicts an age of the universe larger than the one from GCs, it remains on the safe side. However, it is important to emphasize that if a cosmological model is promising in resolving the $H_0$ tension, which directly affects the predicted age, it should not predict an age conflicting with astrophysical estimates such as those from GCs. For instance, the early dark energy (EDE) model, one of the most popular proposals for resolving the $H_0$ tension, typically not only worsens the $S_8$ tension but also predicts the age of the universe to be significantly smaller than the $\Lambda$CDM prediction, even smaller than GC estimates~\cite{Poulin:2023lkg,Bernal:2021yli}. For example, the axion-like EDE, a prominent EDE model, predicts $t_{\rm u} = 13.17^{+0.14}_{-0.15}$ Gyr (Planck+SH0ES), which is in $\sim2\sigma$ tension with the age derived from GCs~\cite{Bernal:2021yli,Boylan-Kolchin:2021fvy,Poulin:2023lkg}. On the other hand, both $\Lambda_{\rm s}$CDM and $\Lambda_{\rm s}$VCDM models predict an age slightly less than $\Lambda$CDM, showing a tension of less than $\sim1\sigma$ (largest $\sim1.2\sigma$ in the case of Planck+BAO) in most of our analyses; see~\cref{tab:results} and~\cref{fig:planck,fig:bao1,fig:bao2}. While the age discrepancies between the $\Lambda$CDM and EDE predictions and GC measurement alone may not seem significant, an astonishing finding in our analyses of both $\Lambda_{\rm s}$CDM and $\Lambda_{\rm s}$VCDM models is that we reduce the $H_0$ tension more, the closer the predicted age of the universe gets to the one from GCs. Notably, in the Planck-alone analysis of the $\Lambda_{\rm s}$VCDM, the predicted $H_0$ and $t_0$ values exhibit no tension at all, only $\sim0.1\sigma$ when considering the SH0ES $H_0$ measurement and the astrophysical age measurement from the oldest GCs.

\begin{figure}[ht!]
    \centering
    \includegraphics[width=9.0cm]{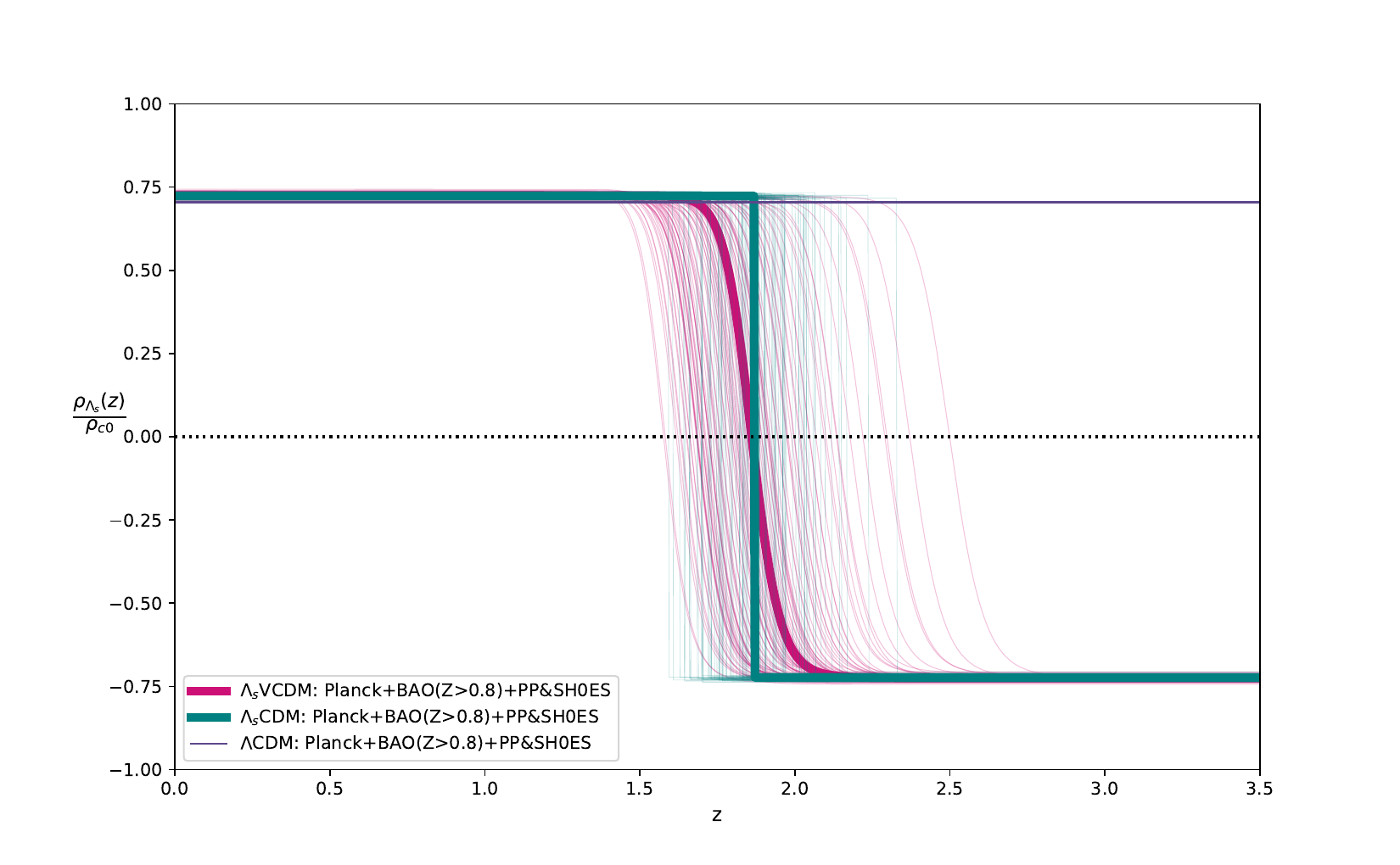}
\includegraphics[width=9.0cm]{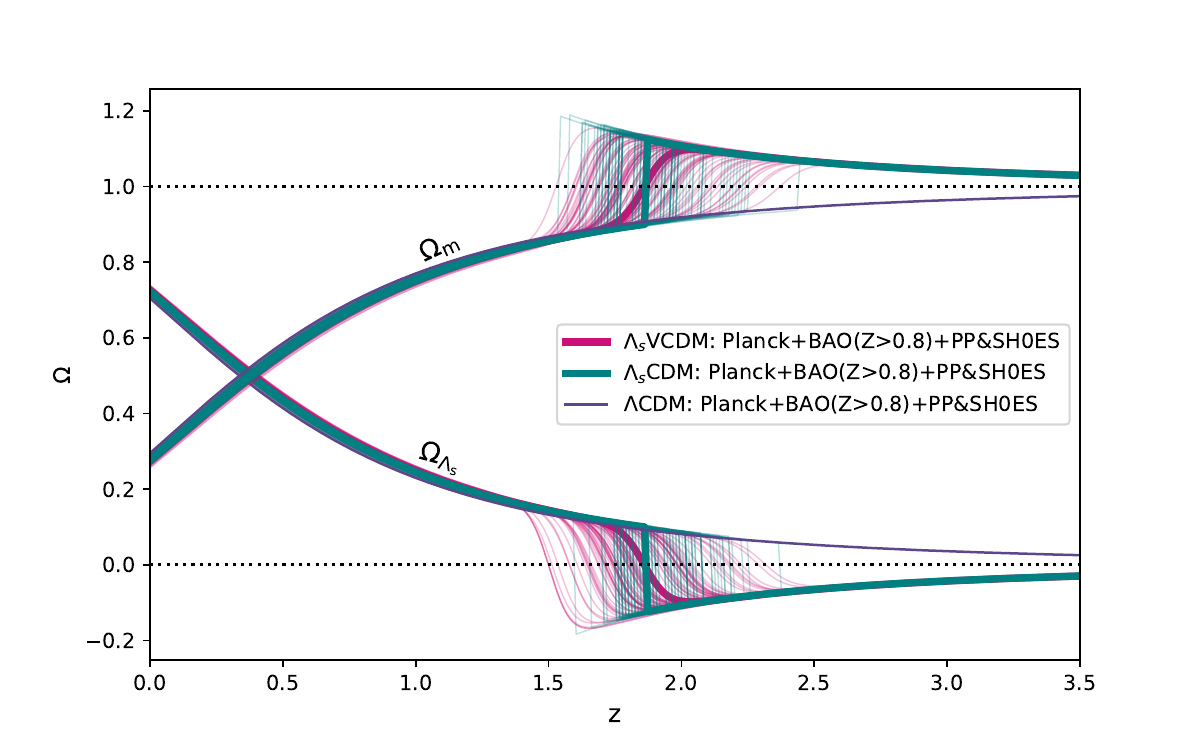}
    \caption{Evolution of the energy density corresponding to $\Lambda$ in the $\Lambda$CDM model and $\Lambda_{\rm s}$ in $\Lambda_{\rm s}$CDM and $\Lambda_{\rm s}$VCDM models, and density parameters ($\Omega$) in redshift ($z$). Upper panel: $\rho_{\Lambda_{\rm s}}(z)/\rho_{c0}$. Lower panel: $\Omega_{\Lambda_{\rm s}}(z) = \rho_{\Lambda_{\rm s}}(z)/\rho_{c}(z)$ and $\Omega_{\rm m}(z) = \rho_{\rm m}(z)/\rho_{c}(z)$. Note the unusual behavior of $\Omega_{\rm m}(z)$; given that $\Omega_{\rm m}(z) \approx 1 - \Omega_{\Lambda_{\rm s}}(z)$, we have $\Omega_{\rm m}(z) > 1$ for $z > z_\dagger$, as for these redshifts, $\Lambda_{\rm s}$ is AdS-like ($\Omega_{\Lambda_{\rm s}} < 0$).}
    \label{fig:d_de}
\end{figure}

Having finalized our main statistical and cosmological interpretations, we aim to quantify the (dis)agreement between the models and the observational data used. To this end, we perform a statistical comparison of the extended models, $\Lambda_{\rm s}$CDM and $\Lambda_{\rm s}$VCDM, with the $\Lambda$CDM model using the Akaike information criterion (AIC) and log-Bayesian evidence, along with $\chi^2_{\text{min}} = -2\ln\mathcal{L}_{\text{max}}$, where $\mathcal{L}_{\text{max}}$ being the maximum likelihood,\footnote{Here, the maximum likelihood is not the mathematical maximum of the likelihood function but rather the maximum likelihood value found in the chains. The same consideration applies to the minimum of $\chi^2$.} as presented in~\cref{tab:results}. Specifically, we first present the relative best fit ($\Delta \chi^2_{\text{min}}=\chi^{2}_{\text{min},\Lambda_{\rm s} \rm (V)CDM}-\chi^{2}_{\text{min},\Lambda \rm CDM}$) and the relative AIC ($\Delta\text{AIC} =\text{AIC}_{\Lambda_{\rm s} \rm (V)CDM}-\text{AIC}_{\Lambda \rm CDM}$, where $\text{AIC} \equiv \chi^2_{\text{min}} + 2N$, with $N$ being the number of free parameters, which serves as the penalization term), both defined with respect to the $\Lambda$CDM model. The preferred model is the one with the smaller AIC value, with negative values of $\Delta\rm AIC$ indicating support for the extended models over $\Lambda$CDM, and more negative values indicating stronger support. By convention, significance of support is judged according to the Jeffreys’ scale, which rates $\Delta\rm AIC>5$ as ``strong" and $\Delta\rm AIC> 10$ as ``decisive" support in favor of the model with the smaller AIC value, regardless of the properties of the models under comparison~\cite{Liddle_AIC}. We also compute the relative log-Bayesian evidence $\Delta \ln\mathcal{Z}\equiv\ln\mathcal{B}_{ij}=\ln \mathcal{Z}_{\Lambda \rm CDM}- \ln \mathcal{Z}_{\Lambda_{\rm s} \rm (V)CDM}$ (where $\mathcal{B}_{ij} =\mathcal{Z}_{i}/\mathcal{Z}_{j}$ is the Bayes' factor with $\mathcal{Z}_i$ and $\mathcal{Z}_j$ being the Bayesian evidences for models $i$ and $j$, respectively) to assess the evidence for the extended models relative to $\Lambda$CDM, using the publicly available package \texttt{MCEvidence}\footnote{\href{https://github.com/yabebalFantaye/MCEvidence}{github.com/yabebalFantaye/MCEvidence}}~\cite{Heavens:2017hkr,Heavens:2017afc}. We follow the convention of assigning a negative value when the extended model, either $\Lambda_{\rm s}$CDM or $\Lambda_{\rm s}$VCDM, is preferred over $\Lambda$CDM, or vice versa. As with the relative AIC, negative values of $\Delta \ln\mathcal{Z}$ imply support for the extended models over $\Lambda$CDM, with more negative values indicating stronger support. We interpret the results using the revised Jeffreys' scale by Trotta~\cite{Kass:1995loi,Trotta:2008qt}; the evidence is classified as inconclusive if $0 \leq | \ln \mathcal{B}_{ij}|  < 1$, weak if $1 \leq | \ln \mathcal{B}_{ij}|  < 2.5$, moderate if $2.5 \leq | \ln \mathcal{B}_{ij}|  < 5$, strong if $5 \leq | \ln \mathcal{B}_{ij}|  < 10$, and very strong if $| \ln \mathcal{B}_{ij} | \geq 10$.

Comparing the $\Lambda_{\rm s}$CDM and $\Lambda_{\rm s}$VCDM models against the standard $\Lambda$CDM model in Table~\ref{tab:results}, we observe that both extended models consistently outperform $\Lambda$CDM, as indicated by negative values of $\Delta \chi^2_{\text{min}}$, $\Delta$AIC, and $\Delta \ln\mathcal{Z}$ across various datasets. These negative values suggest a preference for the extended models over $\Lambda$CDM, with the $\Lambda_{\rm s}$VCDM model generally showing slightly stronger evidence and support compared to $\Lambda_{\rm s}$CDM, though the differences are typically marginal. In particular, $\Delta \chi^2_{\text{min}}$ is negative in all cases, indicating a preference for the extended models. For $\Delta$AIC, the extended models also show negative values, reaffirming this preference, with one exception, though statistically insignificant, in the Planck+BAO dataset for $\Lambda_{\rm s}$CDM, which exhibits a slightly positive value ($\Delta{\rm AIC}=0.7$). For the Planck and Planck+BAO datasets, all three models perform similarly. However, with the Planck+BAO($z>0.8$) dataset, the $\Lambda_{\rm s}$VCDM model receives strong support with $\Delta{\rm AIC}=-7.2$, while the $\Lambda_{\rm s}$CDM model barely achieves strong support with $\Delta{\rm AIC}=-4.7$. When including PP\&SH0ES data, support for both extended models strengthens significantly, reaching decisive levels. Specifically, the preference for both extended models reaches the significance level of approximately $-26$ with the Planck+BAO($z>0.8$)+PP\&SH0ES dataset and around $-14$ with the full data, viz., the Planck+BAO+PP\&SH0ES combination. The $\Lambda_{\rm s}$VCDM model shows stronger preference with a $\Delta{\rm AIC}$ value of $-24.2$, compared to $\Lambda_{\rm s}$CDM, which has a $\Delta{\rm AIC}$ value of $-21.5$ in the case of Planck+PP\&SH0ES. These findings are further supported by Bayesian evidence. In all cases, $\Delta \ln\mathcal{Z}$ is negative, favoring the extended models. Specifically, we find weak evidence for both extended models in the Planck alone analysis. In the case of Planck+BAO, the evidence remains weak for $\Lambda_{\rm s}$VCDM, while it is inconclusive for $\Lambda_{\rm s}$CDM. On the other hand, the evidence strengthens to weak for $\Lambda_{\rm s}$CDM and to moderate for $\Lambda_{\rm s}$VCDM when using the Planck+BAO($z>0.8$) dataset. Including PP\&SH0ES data significantly improves the evidence for both extended models. We find very strong evidence for both extended models with the Planck+PP\&SH0ES and Planck+BAO($z>0.8$)+PP\&SH0ES datasets, and strong evidence for the full data, viz., the Planck+BAO+PP\&SH0ES combination. Notably, the $\Lambda_{\rm s}$VCDM model generally shows slightly stronger evidence and support compared to $\Lambda_{\rm s}$CDM. Thus, from a statistical standpoint, a rapid mirror AdS-dS transition in the late universe, viz., at ${z\sim 1.8-2.2}$, as suggested by the $\Lambda_{\rm s}$VCDM and $\Lambda_{\rm s}$CDM models, performs similarly in explaining the cosmological data and presents a robust alternative to the usual dS-like cosmological constant of the standard $\Lambda$CDM model.

To provide insights into the kinematics of the universe in different scenarios, in \cref{fig:Hz}, we illustrate the comoving Hubble parameter (upper panel), $\dot{a} = H(z)/(1+z)$ (a dot represents a derivative with respect to the cosmological time), Hubble parameter $H(z)$ (middle panel), and the time rate of change of the Hubble parameter (lower panel), $\dot{H}(z)$, scaled by $3H^2(z)$ for the Planck+BAO(${z>0.8}$)+PP\&SH0ES joint analysis. For the $\Lambda_{\rm s}$CDM and $\Lambda_{\rm s}$VCDM scenarios, the panels are produced by doing a weighted sampling from our MCMC chains and plotting for the sampled points using the \href{https://joss.theoj.org/papers/10.21105/joss.00849package}{fgivenx}~\cite{Handley_2018} package; the more frequent the lines, the more probable. 
For the $\Lambda$CDM model, the best-fit prediction from the same joint analysis is shown. Similarly, to provide insights into the dynamics of the sign-switching cosmological constant, $\Lambda_{\rm s}$, particularly regarding the mirror AdS-dS transition epoch, in~\cref{fig:d_de}, we plot the corresponding energy density $\rho_{\Lambda_{\rm s}}(z)$ scaled by the present-day critical energy density $\rho_{\rm c0}=3H^2_0$ and the density parameter for both $\Lambda_{\rm s}$CDM (with an abrupt transition) and the $\Lambda_{\rm s}$VCDM (with a smooth transition) models. \footnote{Note that the variations in the posteriors for $\rho_{\Lambda_s}/\rho_{c0}$ at $z=0$ (corresponding to $\Omega_{\Lambda_{\rm s}}\approx 1 - \Omega_{\rm m}$ at $z=0$), due to the small errors in $\Omega_{\rm m}$, are not clearly visible given the range of the $y$-axis from -1 to 1. A similar situation is observed in the middle and bottom panels of Fig.~5 and the bottom panel of Fig.~6. In contrast, the top panel of Fig.~5, which at $z=0$ corresponds to $H_0$, shows visible variations in the posteriors despite the percent-level constraints on $H_0$, due to the narrower range of the $y$-axis.} For $\Lambda_{\rm s}$CDM, we observe an abrupt (instantaneous) lift in the value of the comoving Hubble parameter by yielding a Dirac-delta distribution at the mirror AdS-dS transition moment. At this transition moment, $\dot{H}$ also exhibits a Dirac-delta distribution (not shown in the plot) resulting in a jump in its value. Namely, $\Lambda_{\rm s}$CDM exhibits a II (sudden) singularity~\cite{Barrow:2004xh} at $z\rightarrow z_\dagger$. One may worry that this can have violent consequences on the structures in the universe; however, a recent work~\cite{Paraskevas:2024ytz} demonstrated that its impact on the formation and evolution of cosmic bound structures is negligible. Therefore, the late time mirror AdS-dS transition does not threaten viability of the $\Lambda_{\rm s}$CDM framework, even in the most extreme case, where an abrupt transition is assumed. For the $\Lambda_{\rm s}$VCDM scenario, we observe a short period of increasing comoving Hubble parameter, indicating a brief period of accelerated expansion ($\ddot{a}>0$). An important point about the $\Lambda_{\rm s}$CDM framework is that the rapid mirror AdS-dS transition does not necessarily imply a period of increasing $H(z)$, i.e., $\dot{H}(z)>0$. Of course, in the abrupt $\Lambda_{\rm s}$CDM scenario, we see an instantaneous jump in the value of $H(z)$ at $z=z_\dagger$.
However, for the $\Lambda_{\rm s}$VCDM model, which features a rapid but smooth transition, the mean value of $\dot{H}(z)$ barely becomes positive, and there is a region of the parameter space from our constraints where it remains always negative. Thus, the $\Lambda_{\rm s}$CDM framework is not characterized by a rapidly (abruptly as a limiting case) increasing late-time epoch of $H(z)$, but rather by a rapid (abrupt as a limiting case) mirror AdS-dS transition in the late universe, around $z=z_\dagger\sim2$, compare the middle panel of~\cref{fig:Hz} with the lower panel of~\cref{fig:d_de}. Additionally, the term \textit{mirror} implies the $\Lambda_{\rm s}$ has the same magnitude before and after the AdS-dS transition, whereas, however, as seen in the lower panel of~\cref{fig:d_de}, the density parameter $\Omega_{\Lambda_{\rm s}}$ abruptly/rapidly assumes negative values for $z>z_\dagger$, but it also rapidly approaches zero.
This explains why the mirror AdS-dS transition most effectively leads to a deviation from the $\Lambda$CDM model if it occurs at a lower redshift. Specifically, the $\Lambda_{\rm s}$CDM and $\Lambda_{\rm s}$VCDM models become indistinguishable from the $\Lambda$CDM model, given the precision of the currently available data, if the transition occurs too early, say, if $z_\dagger\gtrsim4$. At this redshift, whether the cosmological constant is negative or positive, the universe is still highly matter dominated. We see that it is strictly $\Omega_{\rm m}>1$ before the mirror AdS-dS transition begins, which is expected because $\Omega_{\Lambda_{\rm s}}<0$, while $\Omega_{\rm m}+\Omega_{\Lambda_{\rm s}}=1$ as we consider a spatially flat FLRW universe.

We see that there is a parameter space within our constraints where $\dot{H}>0$ for a brief period of time. Within GR, this would imply the violation of the null energy condition (implying $\rho+p\geq0$ for a perfect fluid) by the total energy-momentum tensor of the universe (sign-switching cosmological constant + standard matter fields), signaling the presence of ghosts and/or gradient instabilities.\footnote{Thus, $\dot{H}\leq0$ implies an upper limit on the rapidity of a smooth mirror AdS-dS transition within GR. A detailed investigation of this point is in progress and will be presented in an upcoming paper.} However, in the $\Lambda_{\rm s}$VCDM model~\cite{Akarsu:2024qsi}, realizing $\Lambda_{\rm s}$CDM with a smooth AdS-dS transition in a type II minimally modified gravity called VCDM~\cite{DeFelice:2020eju}, the occurrence of $\dot{H}>0$ is completely safe. All the gravity (i.e., gravitational waves) and standard matter fields remain always stable.\footnote{By stability, we mean the absence of ghosts and/or gradient instabilities, allowing instead the standard Jeans instability for the pressureless components.} It is a crucial property of VCDM that it does not possess extra (scalar or not) degrees of freedom, which, if they existed, could be unstable~\cite{DeFelice:2020eju}, rendering VCDM tailor-made for the $\Lambda_{\rm s}$CDM framework.

\section{Conclusion} \label{sec:conclusions}

We have conducted a study, in light of observational data, focusing on the implementation of $\Lambda_{\rm s}$CDM~\cite{Akarsu:2019hmw, Akarsu:2021fol, Akarsu:2022typ, Akarsu:2023mfb} as a full model by embedding it into the framework of VCDM~\cite{DeFelice:2020eju}, a type II minimally modified gravity theory, as done in a recent work~\cite{Akarsu:2024qsi}. Our primary aim was to determine whether the phenomenology of $\Lambda_{\rm s}$CDM would be compromised after allowing the cosmological perturbations to undergo a period of rapid mirror AdS-dS transition in the late universe. \textit{A priori}, without a specific model, we cannot predict whether the fit to the data will worsen. The embedding of $\Lambda_{\rm s}$CDM into VCDM, giving rise to what we call here $\Lambda_{\rm s}$VCDM, was implemented because the VCDM theory, by construction, possesses only two tensor degrees of freedom in the gravity sector as in general relativity and gives rise to no new scalars in the particle spectrum. This allows for a rapid transition in the dark energy component without leading to any instability. In this paper, we have focused on comparing the fit to the data of three different setups, (i) $\Lambda$CDM, (ii) $\Lambda_{\rm s}$CDM, and (iii) $\Lambda_{\rm s}$VCDM, to critically evaluate and highlight the potential improvements offered by the new model, $\Lambda_{\rm s}$VCDM.

We have shown that the $\Lambda_{\rm s}$CDM paradigm, through both the $\Lambda_{\rm s}$CDM (assuming GR and an abrupt transition) and $\Lambda_{\rm s}$VCDM (assuming VCDM and a smooth transition), successfully addresses the $H_0$ and $S_8$ tensions simultaneously, without causing any inconsistency with astrophysical estimations of the present-day age of the Universe, such as those from the oldest globular clusters.

On the other hand, when comparing these two particular models, namely, $\Lambda_{\rm s}$CDM and $\Lambda_{\rm s}$VCDM, we find differences in the $\chi_{\rm min}^2$ values. $\Lambda_{\rm s}$VCDM presents a lower $\chi_{\rm min}^2$ value for Planck-alone analysis and also when Planck data is combined with BAO/BAO($z>0.8$) or  PP\&SH0ES data, but becomes indistinguishable when all the data are combined. This discrepancy is not surprising, as cosmological observables depend on the combined dynamics of the background and the perturbations, which can differ in GR and VCDM for the same background. In particular, the difference between $\Lambda_{\rm s}$CDM and $\Lambda_{\rm s}$VCDM in this cosmological context can be explained by two key points: (i) the mirror AdS-dS transition, and therefore the transition in the background evolution of the universe, happens abruptly (instantaneously) in the $\Lambda_{\rm s}$CDM model, whereas in the $\Lambda_{\rm s}$VCDM model, the mirror AdS-dS transition is still fast but smoothly extended over a period of time, leading to a new brief temporary accelerated expansion era in the history of the Universe; (ii) the VCDM gravity model distinctly provides the dynamics for the perturbations in the $\Lambda_{\rm s}$VCDM model, including during the transition period, leading to extra terms in the cosmological perturbation equations. These terms are sensitive to the dynamics of the transition (both the background evolution and the dark energy, particularly, while it is undergoing the mirror AdS-dS transition) because they are influenced by the terms proportional to $\Lambda_{{\rm s}, a}$, which is closely related to $\dot H$. In particular, we have managed to constrain the parameter $z_\dagger$ to $z_\dagger=1.88_{-0.58}^{+0.28}$ at 68\% CL even for the Planck-alone analysis of the $\Lambda_{\rm s}$VCDM model, whereas for $\Lambda_{\rm s}$CDM we have only a lower bound of $z_\dagger>1.45$ at 95\% CL. And, notably, the Planck-alone constraint obtained on $H_0$, viz., $H_0=73.4^{+1.8}_{-4.6}~{\rm km\, s^{-1}\, Mpc^{-1}}$, turned out to be in excellent consistency with the latest SH0ES $H_0$ measurements~\cite{Riess:2021jrx,Uddin:2023iob,Breuval:2024lsv}, although we note that this improvement is still partly due to the degeneracy in the $z_\dagger$-$H_0$ plane, it is also significantly reflected in the mean $H_0$ value closely aligning with the SH0ES measurements. This result is nontrivial, as embedding $\Lambda_{\rm s}$CDM into a gravity model could have deteriorated the success of the fit to the data, but in $\Lambda_{\rm s}$VCDM, this does not happen, and some results are even more promising compared to the abrupt $\Lambda_{\rm s}$CDM model. This demonstrates that the conjecture of $\Lambda_{\rm s}$CDM can be successfully implemented into a predictive model, validating its consequences. More work is needed to further understand this improvement in the fit of Planck-alone data by $\Lambda_{\rm s}$VCDM, not only by comparing it to $\Lambda_{\rm s}$CDM but also to $\Lambda$CDM. In particular, we aim to understand the role played by the modified perturbation equations related to the transition and how the background transition itself may change the cosmological observables.

Finally, our findings in this work pave the way for another outcome: although different realizations of the $\Lambda_{\rm s}$CDM framework within the VCDM gravity model, in line with the $\Lambda_{\rm s}$VCDM model, seem to point in an interesting direction, there could be other implementations of $\Lambda_{\rm s}$CDM, i.e., other theory embeddings, that might lead to even better fits to the data. We will pursue these research paths in future projects.

\begin{acknowledgments}
\noindent 
 \"{O}.A. acknowledges the support of the Turkish Academy of Sciences in the scheme of the Outstanding Young Scientist Award (T\"{U}BA-GEB\.{I}P). This study was supported by Scientific and Technological Research Council of Turkey (TUBITAK) under the Grant No.~122F124. The authors thank TUBITAK for their support. The work of A.D.F. was supported by the Japan Society for the Promotion of Science Grants-in-Aid for Scientific Research No.~20K03969. E.D.V. acknowledges support from the Royal Society through a Royal Society Dorothy Hodgkin Research Fellowship.  S.K. gratefully acknowledges the support of Startup Research Grant from Plaksha University  (File No. OOR/PU-SRG/2023-24/08), and Core Research Grant from Science and Engineering Research Board (SERB), Govt. of India (File No.~CRG/2021/004658). R.C.N. thanks the financial support from the Conselho Nacional de Desenvolvimento Cient\'{i}fico e Tecnologico (CNPq, National Council for Scientific and Technological Development) under the project No. 304306/2022-3, and the Funda\c{c}\~{a}o de Amparo \`{a} Pesquisa do Estado do RS (FAPERGS, Research Support Foundation of the State of RS) for partial financial support under the project No. 23/2551-0000848-3. E.\"{O}.~acknowledges the support by The Scientific and Technological Research Council of Turkey (T\"{U}B\.{I}TAK) in the scheme of 2214/A National PhD Scholarship Program. J.A.V. acknowledges the support provided by FOSEC SEP-CONACYT Investigaci\'on B\'asica A1-S-21925, Ciencias de Frontera CONACYT-PRONACES/304001/202 and UNAM-DGAPA-PAPIIT IN117723. A.Y. is supported by a Senior Research Fellowship (CSIR/UGC Ref.\ No.\ 201610145543) from the University Grants Commission, Govt.\ of India. This article/publication is based upon work from COST Action CA21136 – “Addressing observational tensions in cosmology with systematics and fundamental physics (CosmoVerse)”, supported by COST (European Cooperation in Science and Technology).
\end{acknowledgments}

\appendix
\section{THE VCDM MODEL}
\label{sec:app1}

In this appendix, we briefly describe the VCDM model~\cite{DeFelice:2020eju,DeFelice:2020cpt}, a type II minimally modified gravity theory, in which the sign-switching cosmological constant ($\Lambda_{\rm s}$) is embedded~\cite{Akarsu:2024qsi}, setting the foundation for the model referred to as $\Lambda_{\rm s}$VCDM in this paper. In the VCDM model, the usual cosmological constant ($\Lambda$) is replaced by a potential $V(\phi)$ of a nondynamical auxiliary field ($\phi$), avoiding the introduction of extra physical degrees of freedom. The action for this theory is expressed as
\begin{align}
    S &= S_{\rm m} + \Mpl^2 \int {\rm d}^4x \, N \sqrt{\gamma} \left[ \frac{1}{2} \left( R + K_{ij} K^{ij} - K^2 \right) \right. \nonumber \\
    &\left. - V(\phi) + \frac{\lambda_2}{N} \gamma^{ij} D_i D_j \phi - \frac{3\lambda^2}{4} - \lambda (K + \phi) \right] \!,
\end{align}
where $S_{\rm m}$ represents the sum of standard matter actions, $N$ is the lapse function, and $K_{ij}$ is the extrinsic curvature (with $K = \gamma^{ij} K_{ij}$ as its trace) relative to the 3D space metric $\gamma_{ij}$ (which has an inverse $\gamma^{ij}$, determinant $\gamma$, and a covariant derivative $D_i$). The fields $\lambda$, $\lambda_2$, and $\phi$ are auxiliary fields. This modified gravity theory breaks four-dimensional diffeomorphism invariance but retains three-dimensional spatial diffeomorphism invariance and time-reparametrization invariance.

The VCDM theory is fully determined only after the potential $V(\phi)$ is specified, as the contribution of the auxiliary field $\phi$ to the Friedmann equation is given by $\rho_{\phi}\equiv\Mpl^{2}(V-\phi V_{,\phi})+\frac{3}{4}\,\Mpl^{2}\,V_{,\phi}^{2}$, assuming a spatially flat Robertson-Walker metric, as in this work. Alternatively, one can specify the behavior of $\rho_{\phi}(\phi)$ and then determine the corresponding potential. In~\cref{eq:vcdm1}, we have specified the profile for $\Lambda_{\rm s}(a)\equiv\rho_{\phi}(a)/\Mpl^2$, describing an effective smooth sign-switching cosmological constant within the VCDM framework, and then our modified Friedmann equation reads
\begin{equation}
    3\Mpl^2 H^2 = \rho(a) + \Mpl^2 \Lambda_{\rm s}(a)\,,
\end{equation}
where $\rho=\sum_I \rho_I$ is the total matter-energy density, with $I$ running over all the matter components, including the dark matter. Each of these components satisfies the local energy-momentum conservation separately, meaning that each $\rho_I$ is a known function of the scale factor $a$. This implies that $H(a)$ is fully determined. In VCDM, two other equations hold on a spatially flat RW background
\begin{align}
    a\frac{d\phi}{da}&=\frac{3}{2}\,\frac{\rho+P}{\Mpl^2 H}\,,\label{eq:dphi_da}\\
    V&=\frac13\,\phi^2-\frac{\rho}{\Mpl^2}\,,\label{eq:V_phi}
\end{align}
where $P=\sum_I P_I$ is the total pressure for all standard matter components. Since $H=H(a)$ is now a known function of $a$, we can solve~\cref{eq:dphi_da} to find $\phi=\phi(a)$ after fixing the initial condition $\phi(a=1)=-3H_0$ (see \cite{Akarsu:2024qsi}). Then, using~\cref{eq:V_phi}, we obtain the potential $V(\phi)$ (expressed in a parametric form). This process fully determines the $\Lambda_{\rm s}$VCDM theory under consideration in this work. Finally, the modified equation for the perturbations, which can be directly deduced from the Lagrangian, was already given in~\cref{foonote3}.

\bibliography{bib}

\end{document}